\documentclass[useAMS,usenatbib]{mn2e}
\usepackage{graphicx}
\usepackage{natbib}
\usepackage{enumerate}  
\usepackage{amssymb}
\usepackage{longtable,lscape}
\usepackage{rotating}
\usepackage{textcomp}
\usepackage{balance}
\usepackage{placeins}
\usepackage{amsmath}
\usepackage{txfonts}
\usepackage{multicol}
\usepackage[bf]{caption}

\newcommand{\hiiexplorer}{HII{\sc{explorer}}~}
\newcommand{\bpass}{{\small{BPASS}}~}
\newcommand{\cloudy}{{\small{CLOUDY}}~}
\newcommand{\hb}{H$\beta$~}
\newcommand{\ha}{H$\alpha$~}
\newcommand{\hd}{H$\delta$~}

\newcommand{\msun}{M$_{\odot}$~}
\newcommand{\twospace}{$\!\!$}
\newcommand{\onespace}{\hspace{-3pt}}
\newcommand{\starlight}{{\small{STARLIGHT}}~}

\newcommand{\popstar}{{\small{POPSTAR}}~}
\newcommand{\paperone}{\mbox{{Paper~I}}}

\DeclareRobustCommand{\ion}[2]{%
\relax\ifmmode
\ifx\testbx\f@series
{\mathbf{#1\,\mathsc{#2}}}\else
{\mathrm{#1\,\mathsc{#2}}}\fi
\else\textup{#1\,{\mdseries\textsc{#2}}}%
\fi}

\newcommand{\ariv}{[\ion{Ar}{iv}]~}
\newcommand{\ciii}{\ion{C}{iii}~}
\newcommand{\civ}{\ion{C}{iv}~}
\newcommand{\feiii}{\ion{Fe}{iii}~}

\newcommand{\hii}{\ion{H}{ii}~}

\newcommand{\hei}{\ion{He}{i}~}
\newcommand{\heii}{\ion{He}{ii}~}
\newcommand{\nii}{[\ion{N}{ii}]}
\newcommand{\neiv}{[\ion{Ne}{iv}]~}
\newcommand{\niii}{\ion{N}{iii}~}
\newcommand{\niv}{\ion{N}{iv}~}
\newcommand{\nv}{\ion{N}{v}~}

\newcommand{\oii}{[\ion{O}{ii}]}
\newcommand{\oiii}{[\ion{O}{iii}]}
\newcommand{\ovi}{\ion{O}{vi}~}
\newcommand{\sii}{[\ion{S}{ii}]}

\makeatletter
\newcounter{subsubsubsection}[subsubsection]
\renewcommand\thesubsubsubsection{\thesubsubsection .\@arabic\c@subsubsubsection}
\newcommand\subsubsubsection{\@startsection{subsubsubsection}{4}{\z@}%
                                     {-3.25ex\@plus -1ex \@minus -.2ex}%
                                     {1.5ex \@plus .2ex}%
                                     {\normalfont\normalsize}}
\newcommand*\l@subsubsubsection{\@dottedtocline{3}{10.0em}{4.1em}}
\newcommand*{\subsubsubsectionmark}[1]{}
\makeatother

\title[WR population in NGC 3310]{Ionizing stellar population in the disc of NGC 3310 --  II. The Wolf-Rayet population\thanks{Based on observations collected at the Centro Astronómico Hispano-Alemán (CAHA) at Calar Alto, operated jointly by the Max-Planck Institut für Astronomie and the Instituto de Astrofísica de Andalucía (CSIC).}}
\author[Miralles-Caballero et al.]{\parbox{\textwidth}
{D. Miralles-Caballero$^{1}$\thanks{E-mail:
daniel.miralles@uam.es},  F. F. Rosales-Ortega$^{2}$, A. I. D\'iaz$^{1}$, H. Ot\'i-Floranes$^{3,4}$, E. P\'erez-Montero$^{5}$ and S. F. S\'anchez$^{6}$}\vspace{0.4cm}\\
$^{1}$Departamento de F\'isica Te\'orica, Universidad Aut\'onoma de Madrid, 28049 Madrid, Spain\\
$^{2}$Instituto Nacional de Astrof\'isica, \'Optica y Elect\'onica, Luis E. Erro 1, 72840 Tonantzintla, Puebla, Mexico\\
$^{3}$Instituto de Astronom\'ia, Universidad Nacional Aut\'onoma de M\'exico, Apdo. Postal 106, Ensenada B. C. 22800, Mexico\\
$^{4}$Centro de Radioastronom\'\i a y Astrof\'\i sica, UNAM, Campus Morelia, 5809, Mexico\\
$^{5}$Instituto de Astrof\'isica de Andaluc\'ia, CSIC, Apdo. 3004, 18080, Granada, Spain\\
$^{6}$Instituto de Astronom\'\i a, Universidad Nacional Auton\'oma de Mexico, A.P. 70-264, 04510, M\'exico,D.F.}
\begin{document}

\date{Accepted 2014 September 24. Received 2014 September 24; in original form 2014 July 10}

\pagerange{\pageref{firstpage}--\pageref{lastpage}} \pubyear{...}

\maketitle

\label{firstpage}

\begin{abstract}
We use integral field spectroscopy to study in detail the Wolf-Rayet (WR) population in NGC 3310, spatially resolving 18 star-forming knots with typical sizes of 200--300 pc in the disc of the galaxy hosting a substantial population of WRs. The detected emission in the so-called blue bump is attributed mainly to late-type nitrogen WRs (WNL), ranging from a few dozens to several hundreds of stars per region. 
Our estimated WNL/(WNL+O) ratio is comparable to reported empirical relations once the extinction-corrected emission is further corrected by the presence of dust grains inside the nebula that absorb a non-negligible fraction of UV photons. Comparisons of observables with stellar population models show disagreement by factors larger than 2--3. However, if the effects of interacting binaries and/or photon leakage are taken into account, observations and predictions tend to converge. We estimate the binary fraction of the \hii regions hosting WRs to be significant in order to recover the observed X-ray flux, hence proving that the binary channel can be critical when predicting observables. We also explore the connection of the environment with the current hypothesis that WRs can be progenitors to long-duration gamma-ray bursts (GRBs). Galaxy interactions, which can trigger strong episodes of star formation in the central regions, may be a plausible environment where WRs may act as 
progenitors of GRBs. Finally, even though the chemical abundance is generally homogeneous, we also find weak evidence for rapid N pollution by WR stellar winds at scales of $\sim$ 200 pc.

\end{abstract}


\section{Introduction}

Massive stars have a fundamental influence on the interstellar medium (ISM) and galaxy evolution, although their relative number is rather low and their lifetime short. They are responsible for the bulk of the ionization observed in \hii regions in galaxies and enrich the interstellar medium at short time-scales (\mbox{i.e., $<$ 100 Myr}) by returning the nuclear processed material during their lifetime and at their end by going off as supernovae~\citep{Maeder81a,Maeder81b}. They also supply mechanical energy to the ISM via these processes and the ejection of stellar winds on different evolutionary phases~\citep{Freyer03,Freyer06}.

The most massive ($M \geq 25 M_{\odot}$ for $Z_\odot$) and hot O stars evolve to the Wolf-Rayet (WR) phase starting 2--3 Myr after their birth~\citep{Meynet95}. During this phase these stars, that have lost a large part of their H-rich envelope via strong winds~\citep{Maeder90}, can be described as central He-burning cores. They are also considered to be most favoured candidates to one of the most energetic phenomena known today, long duration (\mbox{$\tau > 2$ s}) Gamma-Ray Burst (GRB), produced while they collapse, after supernova explosions, into black holes~\citep{Woosley06,Crowther07}. 

The strong, broad emission lines seen in the spectra of WRs are due to their powerful stellar winds, which are strongly metallicity dependent~\citep{Nugis00}. The wind is sufficiently dense that an optical depth of unity in the continuum arises in the outflowing material. The spectral features are formed far out in the wind and are seen primarily in emission~\citep{Crowther07}. The unique spectroscopic features indicating the presence of WRs, the most common one centred \mbox{at $\sim$ 4680 \AA{}} (AKA blue bump), has permitted their detection individually in Local Group galaxies (e.g.,~\citealt{Massey98,Massey03,Crowther06a,Crowther06b,Neugent12,Hainich14,Sander14}), collectively within knots of local star-forming galaxies (e.g.,~\citealt{Castellanos02,Hadfield06,Kehrig13}) and in single slit/fibre spectra in more distant starbursts (e.g.~\citealt{Guseva00,Perez-Montero07b,Brinchmann08a,Perez-Montero10,Lopez-Sanchez10a,Shirazi12}), and even in moderate redshift GRB hosts~\citep{Han10}. They are also 
significant contributors 
to the average rest-frame UV spectrum of Lyman Break Galaxies~\citep{Shapley03}. The blend of some stellar Carbon lines around \mbox{at $\sim$ 5808 \AA{}}, known as the red bump, represents another characteristic spectral feature of WRs. However, this bump is more difficult to detect than the blue WR bump; in fact, it is almost always weaker~\citep{Guseva00,Fernandes04,Lopez-Sanchez10a}, observed with no detection of the blue bump on a very few occasions in high spatial resolution (\mbox{$\sim$ 10 pc}) studies~\citep{Westmoquette13}.

The investigation of the WR content in galaxies is crucial to test stellar evolutionary models, specially at sub-solar metallicity where more data are needed to constrain the models. Several studies have attempted to reproduce the number of WRs responsible for the observed stellar emission features. Simple calculations involving the \hb emission line combined with the strength of the blue bump already give a hint~\citep{Kunth81}, but more refined theoretical evolutionary models predict that, at a fixed metallicity, the WR to O ratio strongly varies with the age of the starburst~\citep{Cervinyo94,Maeder94,Schaerer98}. Disagreement between observations and models on this ratio, especially at sub-solar metallicities, has induced the appearance of more sophisticated models including the influence of rotation~\citep{Meynet05} and binary evolution~\citep{vanBever03,vanBever07,Eldridge08} in the evolutionary paths followed by massive stars.

During the last few decades many studies focused on large samples of \hii and WR galaxies have been published. While early works relied on narrow-band images of M33 and local low-mass galaxies~\citep{Drissen93a,Drissen93b}, more recent studies have made use of spectra of local and more distant starbursts to search for WRs in giant \hii regions and \hii galaxies. Although the use of spectra is an important advance (the WR features are directly detected), they use single slit/fibre spectra, which can be affected by aperture effects. This can severly affect the comparison of observations with model predictions, as pointed out in~\cite{Kehrig13}, given that the WR population is normally very localized and therefore less extended than the rest of the ionizing stellar population. These studies are also affected by sampling biases (e.g., only the most luminous clusters and \hii-like regions with a very limited range in ionization conditions are selected, or only a single aperture in large sample of galaxies is 
taken). In addition, slit observations may fail in detecting WR features due to their faintness with respect to the stellar continuum emission and their unknown spatial distribution across the galaxy. Using these techniques, several studies on spatially resolved WR population haven been accomplished for nearby galaxies outside the Local Group (e.g.,~\citealt{Drissen99,Hadfield06,Bibby10,Karthick14}). However, works on resolved WR population using relatively recent techniques that help mitigate the mentioned biases (e.g.,~\citealt{Bastian06,Monreal10,Kehrig13}) are not numerous. With the advent of the integral field spectrospy (IFS), these limitations can be overcome or at least significantly diminished. IFS has proved to be a powerful technique in minimizing the WR bump dilution and finding WRs in extragalactic systems where they were not detected before~\citep{Kehrig08,Cairos10,Garcia-Benito10}. With this powerful technique we can have simultaneous spectral and spatial information, which can allow 
us to study in a more efficient way the spatial distribution of the WR population in local galaxies.

The PPAK Integral-field-spectroscopy Nearby Galaxies Survey (PINGS;~\citealt{Rosales-Ortega10}) is a survey specially designed to obtain complete maps of the emission-line abundances, stellar populations and extinction using an IFS mosaicking imaging for nearby (\mbox{$D_\mathrm{L} \leq $ 100 Mpc}) 17 well-resolved spiral galaxies. In~\cite{Miralles-Caballero14}, hereafter \mbox{Paper~I},  we  characterized the properties (i.e., age, mass, impact on gas metallicity) of the ionizing population in 99 identified \hii regions in the stellar disc of NGC 3310 in order to study the impact of the minor merger on the star formation properties of the remnant. This moderately-low metal galaxy (\mbox{12 + log(O/H) $\sim$ 8.2--8.4};\citealt{Pastoriza93}; \mbox{Paper I}) is a very distorted spiral galaxy classified as an SAB(r)bc by~\cite{deVaucouleurs91}, with strong star formation. With a broad range of stellar masses from about 10$^4$ to 6$\times 10^{6}$ \msun and a narrow age range of about 2.5-5 Myr, WR 
signatures are expected to be observed at 
least for the most massive population. These features were early mentioned in the intense star-forming region known as Jumbo in~\cite{Pastoriza93}, and in several \hii regions in \mbox{Paper~I}. In the present study, we have identified a sample of 18 \hii regions with measurable WR features in their spectra distributed across the disc of NGC 3310. The main strength of our data resides in the knowledge of the spatial location of the WR population, so that we can characterize it and investigate how it affects the environment. In particular, we analyse the WR content within these \hii regions, investigate if current synthesis stellar models are able to explain their emission properties and if WR are able to rapidly (\mbox{i.e.,~$\tau \sim 10$ Myr}) enrich the interstellar medium at spatial scales close to 200 pc. The importance of including binaries in models is additionally explored with the aid of X-ray spectra, which help us constrain the binary fraction of the stellar population. The integrated properties (
i.e., luminosity-metal and mass-metal relations) of the galaxy are also compared with those of more distant GRB hosts that present WR features to look for hints about the driving mechanisms of these energetic phenomena, taking into account that, as mentioned above, WRs are widely assumed to be progenitors of long duration GRBs.   

The paper is organized as follows. We briefly present the dataset used in Sect.~\ref{sec:obs}. The identification of \hii regions with WR features and the procedure used to fit the broad and nebular emission lines in the blue bump are explained in Sect.~\ref{sec:ananlysis_res}. The derived content of WRs and ratios to O stars are also reported in this section. In Sect.~\ref{sec:discussion} we compare the observables with the predictions of synthesis stellar models, investigate on the binarity of the stellar population, speculate on the WR-GRB connection and discuss on the metal enrichment due to WR star winds. Finally, we compile our main conclusions in Sect.~\ref{sec:conclusions}. Throughout this paper the luminosity distance to NGC 3310 is assumed to be 16.1 Mpc (taken from the NASA Extragalactic Database; NED). With an adopted Cosmology of \mbox{H$_0$ = 73 km s$^{-1}$ Mpc$^{-1}$} an angle of 1\arcsec~corresponds to a linear size of 78 pc.

\section[]{Observational data}
\label{sec:obs}
\subsection{PINGS data}

NGC 3310 observations were carried out with the 3.5m telescope of the Calar Alto observatory using the Postdam Multi-Aperture Spectrograph (PMAS;~ \citealt{Roth05}) in the PMAS fibre Package mode (PPAK;~\citealt{Verheijen04},~\citealt{Kelz06}). This was part of the PPAK IFS Nearby Galaxies Survey (PINGS;~\citealt{Rosales-Ortega10}). The V300 grating was used to cover the 3700-7100 \AA{} spectral range with a spectral resolution (FWHM) of \mbox{10 \AA{}}, corresponding to \mbox{600 km s$^{-1}$}. Three pointings with a dithered pattern (3 dithered exposure per pointing) were taken, observing strategy that allowed us to re-sample the PPAK 2.7\arcsec-diameter fibre to a final mosaic with a 1-arcsec spaxel and a field of view (FoV) of about \mbox{148 $\times$ 130 arcsec$^2$}. The seeing of the observations was on average 1.5\arcsec. We estimated the angular resolution to be about the size of the fibre (i.e., spatial resolution of about 200 pc) by using a rather isolated foreground star within the FoV for 
NGC 1057 (there is none in the FoV of NGC 3310), which was observed within the PPAK programme during the same night as NGC 3310, and thus the observation conditions were practically the same. 

The technical specifications of the observations, data reduction (which followed the standard procedures for this kind of data as described in~\citealt{Sanchez06}) and absolute flux calibration are summarized in \paperone. The reduced IFS data were stored in a 3-dimensional FITS image, with two spatial dimensions and one corresponding to the dispersion axis. A total of 8705 spectra were finally produced, spatially resolved in spaxels of \mbox{1 $\times$ 1 arcsec$^2$}. 

\subsection{X-ray data}

We obtained \textit{XMM-Newton} data for NGC 3310 (observations ID 0556280101 and 0556280201). The observations were taken in 2008 and 2009, with the two MOS and the pn cameras that comprise the European Photon Imaging Camera (EPIC). We did not analyse available data (ID 010112810301) from observations taken in 2001 because there was a high level of flaring throughout almost the entire observation~\citep{Jenkins04}.

We reduced the observation data files (ODF) using {\small SAS}\footnote{http://xmm.esac.esa.int/sas/} version 13.5.0. The {\small SAS} \textit{epproc} and \textit{emproc} tasks were used to generate the calibrated events files from the raw EPIC data. For each observation, the net useful integration times were of the order of 15 ks for the pn camera and from 36 to 41 ks for the MOS cameras. From these data X-ray images in the energy interval \mbox{2--10 KeV} were created. We then extracted in each case the spectrum of the central region using a circular aperture of 35\arcsec~in radius (see Fig.~\ref{fig:detection_im}). The background was estimated from a region close to the source in the same CCD and free of any contaminating source. The background region was somewhat larger than the aperture used for extracting the spectrum of the galaxy. The energy redistribution matrices were generated with \textit{rmfgen} and \textit{arfgen}. We re-binned all the spectra in order to have at least 25 counts in each 
spectral bin. 

We also retrieved data taken for NGC 3310 with the Advanced CCD Imaging Spectrometer from the \textit{Chandra} archive. The Chandra sequence number of this observation is 600276, and the observation ID 2939. The total exposure time for this observation corresponds to 47.2 ks. We used the type 2 event file  provided by the standard pipeline processing. The CIAO\footnote{http://cxc.harvard.edu/ciao/} package, version 4.5, was used in order to obtain X-ray images with a superb angular resolution of about 0.5\arcsec~within the energy range 2--8 keV. 

\section[]{Analysis and results}
\label{sec:ananlysis_res}
\subsection{Selection of \hii regions with WR features}

\begin{figure}
\centering
\includegraphics[angle=90,trim = 0.5cm 4cm 0cm 3cm,clip=true,width=0.95\columnwidth]{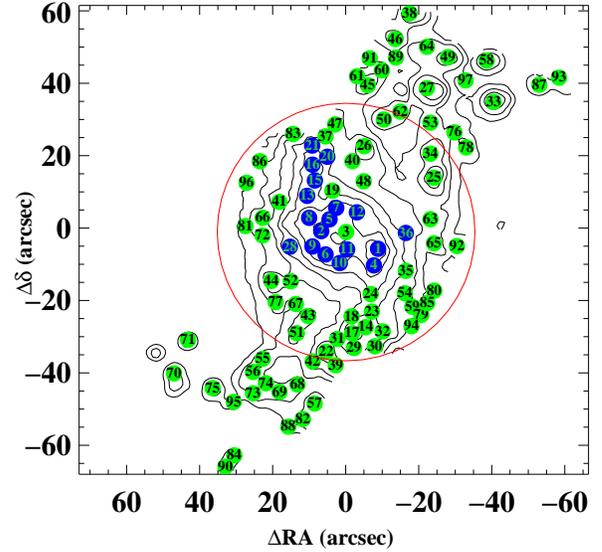}
  \caption{\ha contour map showing the spatial 2D distribution of the identified \hii regions in NGC 3310. Regions with clear WR features are coloured in blue. The rest of the sample is represented as circles coloured in green. The large red circle shows the approximate aperture used to extract the X-ray spectra.}
  \label{fig:detection_im}
\end{figure}

The main Wolf-Rayet features seen in the optical spectra are two broad emission features: the blue bump around 4600--4680 \AA{} and the red bump around 5670--5820 \AA{}. In \paperone~ a total of 99 \hii regions were identified using the semi-automatic procedure \hiiexplorer (\citealt{Sanchez12b}) on the \ha images. As a result, we obtained a spectrum for each \hii region. We have thus searched for those characteristic features in our spectra.
As explained in \paperone, the stellar continuum was subtracted from each of the rest-frame observed spectrum using the spectral synthesis code \starlight 
(\citealt{Cid-Fernandes04b,Cid-Fernandes05}). The residual spectrum is supposed to be formed by only nebular emission lines. However, the WR stellar emission features are not implemented in the  models. Thus, these features remain in this residual spectrum, which is the one used to perform our study. Note that the nebular continuum was also subtracted and therefore the continuum of this spectrum is theoretically flat and of zero value. 

We have followed a simple procedure to select those \hii regions with ``detectable'' WR features: (i) the rms is computed in a local spectral range close to the WR features; (ii) we then compute the peak value of the spectrum within the spectral range that the WR features cover; (iii) those \hii regions with a flux ratio \mbox{$ F_{\mathrm{peak}} / \mathrm{rms} > 5$} are selected; (iv) finally, a visual inspection is performed on the selected regions given that in a few cases the peak emission corresponds to a single nebular line or to an artefact. Following this procedure, a sample of 18 \hii regions with clear evidence of a blue bump was selected, being ID 20 the region with the faintest detectable WR feature. As Fig.~\ref{fig:detection_im} illustrates, they are distributed in the central region of the galaxy and along the spiral northern arm. Interestingly, we did not find clear WR features in other more external regions (regions ID 14, 17, 22, 27 and 45) with similar estimated age and mass (see \paperone)
 and lower observed stellar contiuum than in ID 20. Unfortunately, we did not find a clear evidence of the red bump emission in any spectrum, as Fig.~\ref{fig:red_bump} shows.

Other weak stellar optical lines, originated in WRs and that can give us an idea of what subtype of WR is present, can also be observed. We also searched for them when possible. As Fig.~\ref{fig:red_bump} illustrates, the detection of \mbox{\ciii 5696 \AA{}} was not positive for any of the regions. Given the spectral resolution of our data (\mbox{FWHM $\sim$ 10 \AA{})}, we can not resolve the emission of \mbox{\niii 4097 \AA{}} and \mbox{\ovi 3834 \AA{}}, since they lie too close to \hd (at 4101 \AA{}) and H8 (at 3835 \AA{}), respectively. Finally, we did not find any clear evidence of \mbox{\ovi 3811 \AA{}}.

\subsection{Identification of the WR features within the bump}

\begin{figure*}
\centering
\includegraphics[trim = 0cm 0cm 0cm 4cm,clip=true,width=0.90\textwidth]{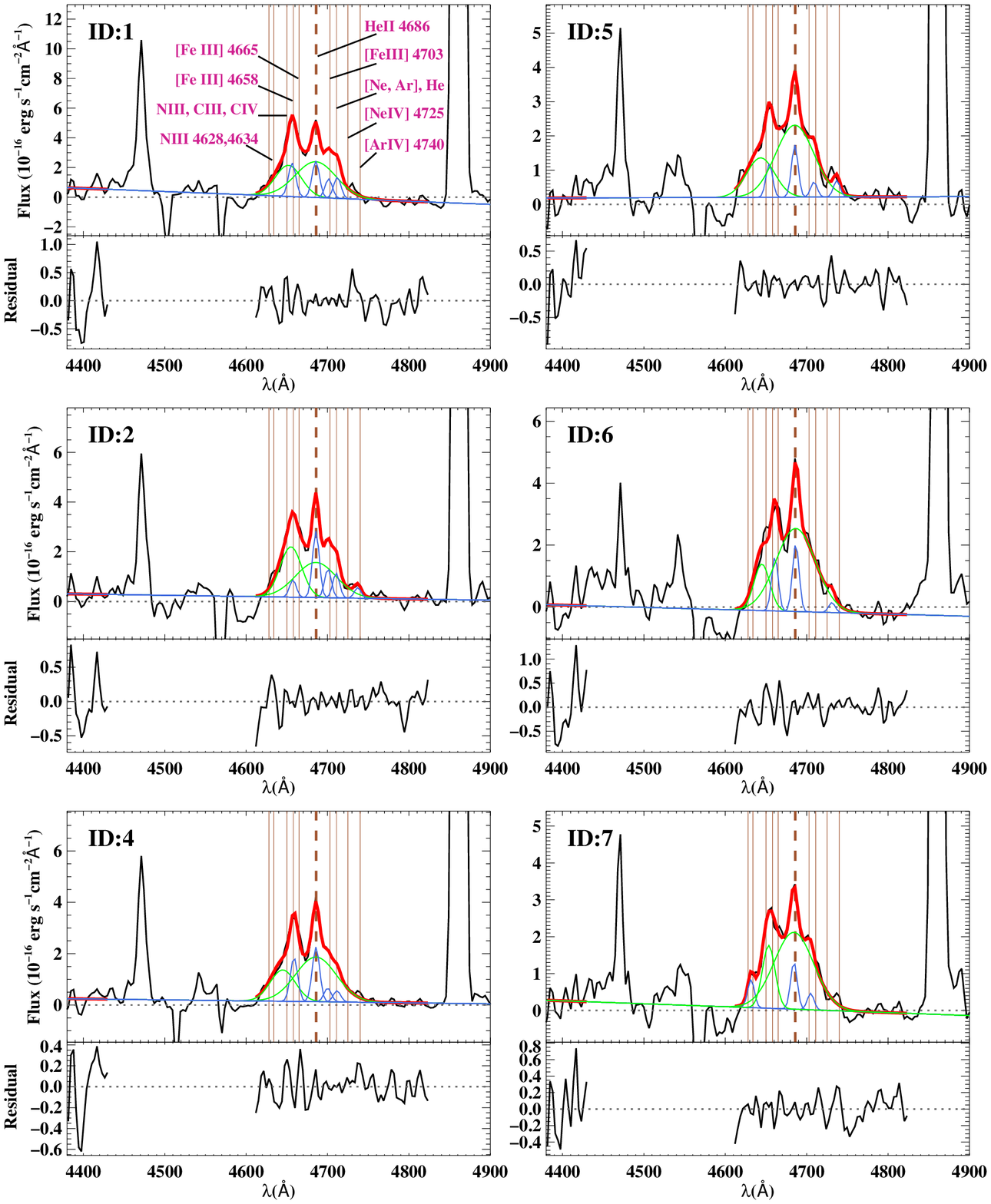}
  \caption{Multiple line-fit of WR features within the blue bump. In each figure the \textit{residual} spectrum is shown in black. This corresponds to the emission gaseous and stellar-line spectrum (obtained in \mbox{Paper I} as a result of the \starlight fitting) minus the modelled feature, in flux units. An almost horizontal blue line denotes the resulting continuum of the fit. The total fitted continuum + emission lines on the blue bump is drawn in thick-red line. Nebular (blue) and broad stellar (green) components of the fit are also drawn. Finally, vertical brown lines indicate the position of the nebular and stellar blend lines typically observed in this spectral range when emission from WRs is detected. They are labelled in the figure that corresponds to region ID 1.}
  \label{fig:wrb_fits}
\end{figure*}

\begin{figure*}
 \includegraphics[trim = 0cm 0cm 0cm 4cm,clip=true,width=0.95\textwidth]{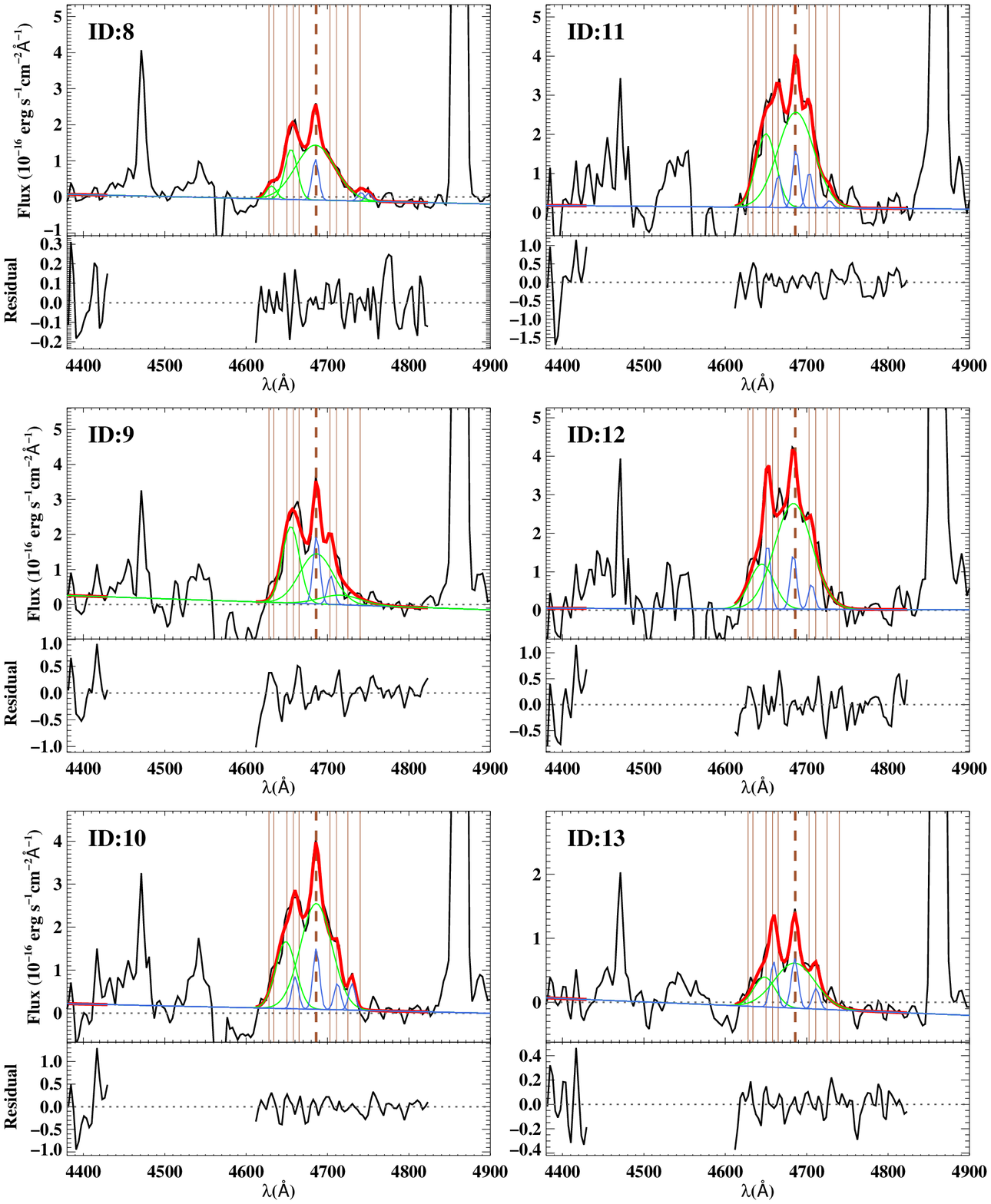}
\addtocounter{figure}{-1}   
    \caption{-- Continued}
\end{figure*}

\begin{figure*}
 \includegraphics[trim = 0cm 0cm 0cm 5cm,clip=true,width=0.95\textwidth]{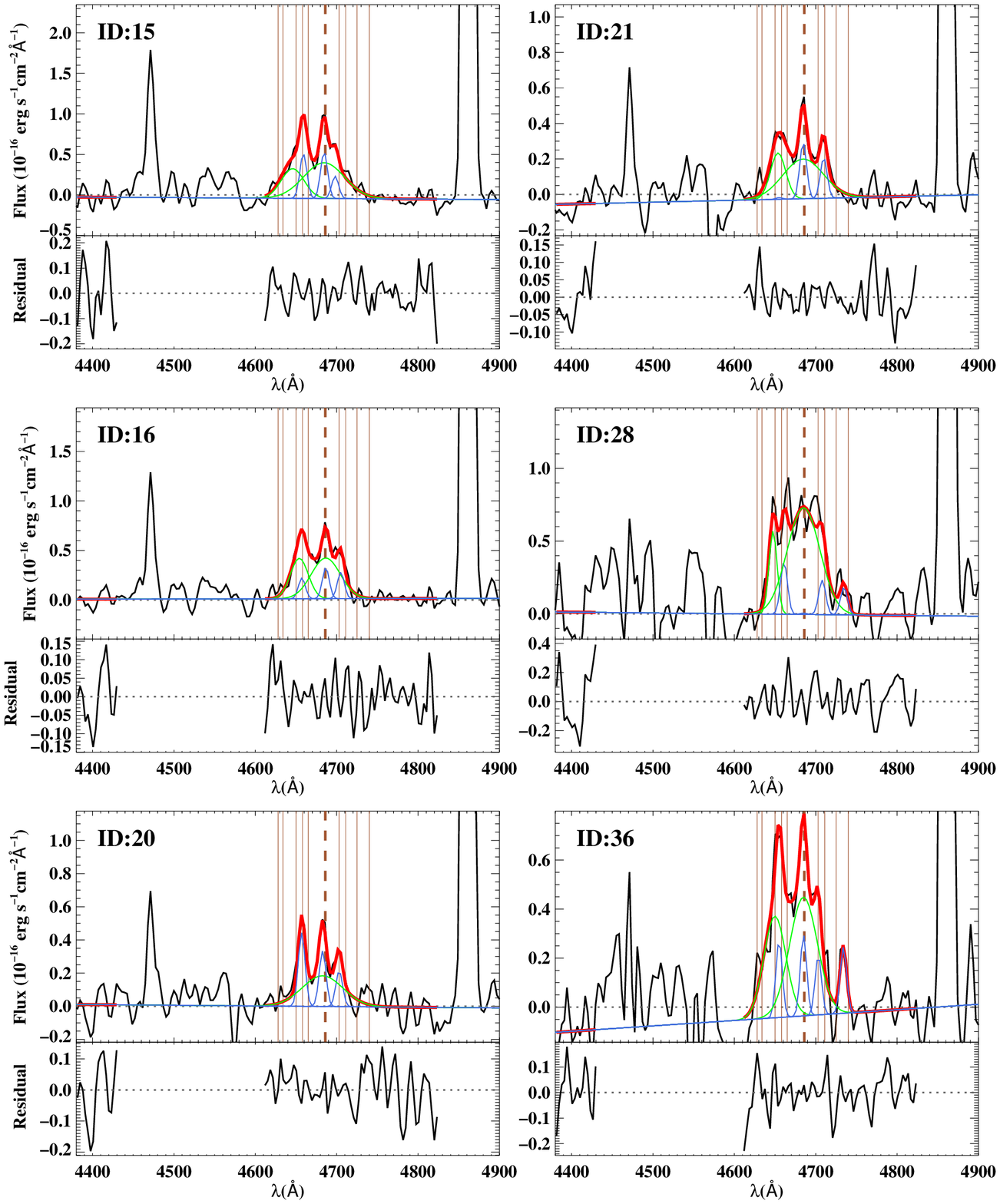}
\addtocounter{figure}{-1}   
    \caption{-- Continued}
\end{figure*}

The so-called blue bump is actually the blend of some broad WR lines, such as the \nv 4605,4620 \AA{}, \niii 4628,4634,4640 \AA{}, \ciii\twospace/\civ 4650, 4658 \AA{} and \heii 4686 \AA{}~ (e.g., \citealt{Conti89,Guseva00,Crowther07}, and references therein). In addition to this, some nebular emission lines, such as [\feiii\onespace] 4658,4665,4703 \AA{}, \ariv 4711,4740 \AA{}, \mbox{\hei 4713 \AA{}} and \mbox{\neiv 4714,4725 \AA{}}, are usually superposed on the bump~\citep{Izotov98,Guseva00}.  Therefore, particular care has to be taken in order to fit the WR features in the blue bump, and considering just one single Gaussian fit is not a good choice, as pointed out by~\cite{Brinchmann08a}. Furthermore, the nebular emission lines within the bump should be properly removed and not included in the flux of the broad stellar lines.

\begin{table*}
\begin{minipage}{0.9\textwidth}
\renewcommand{\footnoterule}{}  
\begin{normalsize}
\caption{Emission measurements of WR features for \hii regions with a clear signature of a WR population in their spectra.}
\label{table:wr_flux_ew}
\begin{center}
\begin{tabular}{lcccccccc}
\hline \hline
   \noalign{\smallskip}
\multicolumn{1}{l}{HII} & 
\multicolumn{1}{c}{H$\beta$} & 
\multicolumn{1}{c}{c (H$\beta$)\footnote{As reported in table A.1 in \paperone}} \vline & 
\multicolumn{2}{c}{Narrow \heii  $\lambda$4686} \vline & 
\multicolumn{2}{c}{Blue WR bump} \vline & 
\multicolumn{2}{c}{\niii / \ciii / \civ~~$\lambda$4650} \\

\multicolumn{1}{l}{ID} & 
\multicolumn{1}{c}{($\times$ 10$^{-14}$ cgs)} &
\multicolumn{1}{c}{} \vline &
\multicolumn{1}{c}{I / I(H$\beta$)} & 
\multicolumn{1}{c}{EW (\AA{})} \vline & 
\multicolumn{1}{c}{I / I(H$\beta$)} & 
\multicolumn{1}{c}{EW (\AA{})} \vline & 
\multicolumn{1}{c}{I / I(H$\beta$)} &
\multicolumn{1}{c}{EW (\AA{})} \\

 \hline
   \noalign{\smallskip}
1& 41.67 $\pm$ 0.20& 0.13 $\pm$ 0.03& 0.96 $\pm$ 0.17& 0.67 $\pm$ 0.17& 5.2 $\pm$ 1.5& 3.7 $\pm$ 1.0& 2.92$\pm$ 0.65& 2.04$\pm$ 0.60\\
2& 26.10 $\pm$ 0.12& 0.16 $\pm$ 0.03& 1.67 $\pm$ 0.18& 1.06 $\pm$ 0.21& 5.3 $\pm$ 1.2& 3.4 $\pm$ 0.8& 4.05$\pm$ 0.46& 2.56$\pm$ 0.52\\
4& 20.66 $\pm$ 0.11& 0.07 $\pm$ 0.03& 1.43 $\pm$ 0.15& 1.01 $\pm$ 0.16& 6.5 $\pm$ 0.7& 4.6 $\pm$ 0.7& 2.81$\pm$ 0.33& 1.98$\pm$ 0.35\\
5& 15.45 $\pm$ 0.09& 0.08 $\pm$ 0.03& 1.28 $\pm$ 0.23& 0.60 $\pm$ 0.17& 10.2 $\pm$ 1.4& 4.8 $\pm$ 1.0& 4.11$\pm$ 0.79& 1.94$\pm$ 0.57\\
6& 15.40 $\pm$ 0.08& 0.05 $\pm$ 0.03& 1.65 $\pm$ 0.28& 0.57 $\pm$ 0.12& 12.0 $\pm$ 1.3& 4.1 $\pm$ 0.6& 2.70$\pm$ 0.62& 0.92$\pm$ 0.25\\
7& 14.69 $\pm$ 0.07& 0.09 $\pm$ 0.03& 1.16 $\pm$ 0.22& 0.52 $\pm$ 0.12& 11.2 $\pm$ 1.2& 5.0 $\pm$ 0.7& 2.95$\pm$ 0.40& 1.32$\pm$ 0.25\\
8& 24.60 $\pm$ 0.12& 0.28 $\pm$ 0.02& 1.05 $\pm$ 0.12& 0.66 $\pm$ 0.13& 8.2 $\pm$ 1.0& 5.2 $\pm$ 0.7& 2.43$\pm$ 0.23& 1.52$\pm$ 0.27\\
9& 17.64 $\pm$ 0.08& 0.14 $\pm$ 0.03& 1.71 $\pm$ 0.32& 0.71 $\pm$ 0.17& 6.2 $\pm$ 4.1& 2.6 $\pm$ 1.3& 4.87$\pm$ 1.32& 2.00$\pm$ 0.66\\
10& 9.82 $\pm$ 0.05& 0.00 $\pm$ 0.03& 1.45 $\pm$ 0.40& 0.46 $\pm$ 0.14& 12.1 $\pm$ 1.8& 3.8 $\pm$ 0.7& 4.89$\pm$ 1.25& 1.54$\pm$ 0.45\\
11& 12.22 $\pm$ 0.08& 0.05 $\pm$ 0.03& 1.39 $\pm$ 0.55& 0.42 $\pm$ 0.19& 12.7 $\pm$ 2.5& 3.8 $\pm$ 0.9& 5.38$\pm$ 1.55& 1.63$\pm$ 0.57\\
12& 10.90 $\pm$ 0.07& 0.00 $\pm$ 0.01& 1.26 $\pm$ 0.34& 0.36 $\pm$ 0.13& 14.5 $\pm$ 1.5& 4.1 $\pm$ 0.8& 4.05$\pm$ 0.99& 1.16$\pm$ 0.38\\
13& 10.03 $\pm$ 0.04& 0.15 $\pm$ 0.03& 1.26 $\pm$ 0.24& 0.66 $\pm$ 0.16& 6.8 $\pm$ 1.4& 3.5 $\pm$ 0.7& 2.39$\pm$ 0.75& 1.24$\pm$ 0.48\\
15& 5.93 $\pm$ 0.03& 0.05 $\pm$ 0.03& 1.18 $\pm$ 0.19& 0.91 $\pm$ 0.21& 5.3 $\pm$ 0.9& 4.1 $\pm$ 0.9& 2.47$\pm$ 0.59& 1.89$\pm$ 0.59\\
16& 5.64 $\pm$ 0.03& 0.10 $\pm$ 0.03& 0.80 $\pm$ 0.23& 0.59 $\pm$ 0.21& 4.4 $\pm$ 1.0& 3.2 $\pm$ 0.7& 2.68$\pm$ 0.48& 1.97$\pm$ 0.48\\
20& 2.53 $\pm$ 0.02& 0.00& 1.39 $\pm$ 0.30& 1.01 $\pm$ 0.29& 4.6 $\pm$ 1.0& 3.3 $\pm$ 1.0& \ldots & \ldots \\
21& 3.30 $\pm$ 0.02& 0.12 $\pm$ 0.03& 1.43 $\pm$ 0.27& 0.76 $\pm$ 0.19& 6.0 $\pm$ 1.4& 3.2 $\pm$ 0.7& 2.67$\pm$ 0.51& 1.41$\pm$ 0.35\\
28& 3.00 $\pm$ 0.02& 0.00 $\pm$ 0.02& \ldots & \ldots & 12.3 $\pm$ 2.1& 4.4 $\pm$ 1.1& \ldots & \ldots \\
36& 1.74 $\pm$ 0.01& 0.00 $\pm$ 0.01& 1.89 $\pm$ 0.64& 0.39 $\pm$ 0.16& 12.0 $\pm$ 2.3& 2.4 $\pm$ 0.6& 8.11$\pm$ 1.60& 1.66$\pm$ 0.45\\
\hline \noalign{\smallskip}
\multicolumn{9}{@{} p{\textwidth} @{}}{\textbf{Notes.} The intensity of the WR features are given relative to the intensity of \hb, normalized to I(\hb\onespace) = 100. No correction for absorbed UV photons in the nebula has been applied for the values given in the table. The EWs have been corrected by the presence of an underlying non-ionizing population.}
\end{tabular}
\end{center}
\end{normalsize}
\end{minipage}
\end{table*}

In Fig.~\ref{fig:wrb_fits} we show a zoomed view of the rest-frame blue bump whereabouts in the continuum extracted spectra as a result of the \starlight fitting  for our sample of 18 \hii regions with clear WR features. The \heii 4686 \AA{} broad feature is easily identifiable in the spectra, together with narrow peaks centred usually at 4650 and 4686 \AA{}. Depending on the quality of the spectra, other narrow lines are also visible. We then first tried to fit the easily identifiable broad + narrow components using Gaussians. The adjacent continuum (expected to be around 0 in these spectra), up to a wavelength of 4830 \AA{} for the right part and spanning the interval
\mbox{4380--4430\AA{}} for the left part, was simultaneously fitted with a one-degree polynomial function. We did not include in the fit the \mbox{4500--4600} \AA{} window since in most cases strong absorption features, probably residuals of badly subtracted background features, are present. The width ($\sigma$) of the narrow lines (i.e., nebular) was fixed to that of \hb while the width of the broad line (i.e., the bump itself) was set free with a maximum value of \mbox{25 \AA{}} \mbox{(FWHM = 59\AA{})}. For the broad component this corresponds to a velocity dispersion of about \mbox{1600 km s$^{-1}$}, an adequate upper-limit to the width of individual WR features (see~\citealt{Smith82,Crowther07,Brinchmann08a}). It became evident that the addition of another broad component at $\sim$ 4650 \AA{} was needed, representing the contribution of the \niii and \ciii/\civ line blends. We iteratively fitted these features with broad and narrow Gaussians and examined the residuals. During this process, if 
the peak of the residual was found larger than \mbox{4$\times$ rms} a new narrow component (e.g., [\feiii] 4665, 4703 ...) was added to the fitting procedure. 

In general, between five and six components in total are needed to properly fit the WR features. Relatively high excitation nebular lines (i.e., \mbox{[\niv\onespace] 4725 \AA{}} and \ariv 4740) were detected in a few regions (ID 2, 5, 8, 10, 21, 36), although the derived uncertainties are of the order of 40--60\%. Table~\ref{table:wr_flux_ew} compiles the results of our fits for the narrow and broad components of \heii $\lambda$4686, and for the broad \niii and \ciii/\civ blends. In particular, the flux intensity (corrected for internal extinction) relative \hb and the EW are provided. Note that the EWs are corrected for the presence of underlying non-ionizing populations, as estimated in \mbox{Paper~I}. To summarize, the star formation history is derived using \starlight. Then, to estimate the continuum emission of the ionizing population, the flux of the non-ionizing population (given the uncertainties, it is assumed as the emission of the modelled stellar population older than 15 Myr) is 
subtracted from the observed spectrum. The typical relative uncertainties in the measurement of these features are larger than 10\% and up to about 40\% (specially for the case of the EW determinations), which is not unexpected given 
that these features are usually weak.  

\subsection{Nature and number of WRs}
\label{sec:number_wr}

The knowledge of the nature of the WR population and the estimation of their number is not straightforward. The \mbox{\heii 4686 \AA{}} emission line and the blue WR bump are mainly linked to WN stars. The absence of \mbox{\niii 4097} and the \mbox{\nv 4605,4620 \AA{}} blend suggests that the dominant population of WN stars corresponds to late-types (WNL), although some contribution by early-type WN stars (WNE) might be still present in the blue WR bump (\citealt{Schaerer98}). In fact,~\cite{Pastoriza93} actually managed to barely identify the \nv blend in a higher spectral resolution single-slit spectrum of the Jumbo (regions ID 1 and 4). The \mbox{\civ 5801,5812 \AA{}} emission feature (the red WR bump) essentially originates in WC stars (mainly in early-types, WCE). However, this feature is usually weaker than the blue bump and its non-detection does not necessarily imply the absence of WC stars. Other lines, directly linked with other WR subtypes (e.g., \ciii 5696 \AA{}, mainly from WCL types; \ovi 3811, 
3834, from WO types) are not detected either. Both WN and WC stars contribute to the emission of the broad \niii 4634,4640 \AA{} (WN), \mbox{\ciii/\civ 4650,4658 \AA{}} (WC) blends. 

If the dominant WN population corresponds to WNL types we can assume that there are no WNE present in order to obtain a first order estimation of the number of WNL using the blue bump, in particular the \mbox{\heii 4686 \AA{}} broad emission feature. Three further considerations have to be taken into account:

\begin{itemize}
 \item In very young bursts (with age \mbox{$\tau <$ 3 Myr}), the \heii ionizing continuum is dominated by O stars~\citep{Schaerer98}. \heii lines are also produced by OIf stars, emitting about \mbox{2$\times 10^{35}$ erg s$^{-1}$} each such star (\citealt{Schaerer98,Brinchmann08b}). At high metallicity these do not contribute greatly to the \heii line luminosity (just a few \% around solar metallicity). However, at low metallicity the WR phase is less prominent and the emission in the blue bump produced by OIf stars can even dominate during the first 2-3 Myr of the burst.  
 
 \item The luminosity of WNL stars is not constant and can vary within factors of a few. Table~\ref{table:wnl_lum} lists several references with adopted values in different observational and theoretical studies, most of them based on observations of WRs in the Galaxy and the Small and Large Magellanic Clouds. In fact, there is increasing evidence that the WR line luminosities are lower at lower metallicities as a consequence of the WR winds being metallicity dependent (\citealt{Crowther06a,Crowther06b,Crowther07}). The uncertainties of the reported values can easily be larger than 50\% due to the poor statistics of the average values, which include only a bunch of observed WRs. The strong dependence of the WR features on the physical processes involving stellar mass loss is also responsible for the high dispersion.
 
 \item Actually, all WR types contribute to the \heii broad emission. According to~\cite{Schaerer98}, in WCL stars, \mbox{\heii 4686 \AA{}} contributes on  average $\sim$12\% to the combined 4650/4686 blend. Likewise, they assume that \mbox{\heii 4686  \AA{}} contributes 8-30\% of the same combined blend for WCE in the LMC, being on average also about 12\%~\citep{Smith90}. The average emission feature of a WC4 star (i.e., WCE) in the combined blend is about \mbox{50$\times 10^{35}$ erg s$^{-1}$}~\citep{Schaerer98,Crowther06a}; for a WCL star is about  \mbox{10$\times 10^{35}$ erg s$^{-1}$}~\citep{Schaerer98}. We can then estimate the emission of these stars in \mbox{\heii  4686  \AA{}} to be about \mbox{1--6 $\times 10^{35}$ erg s$^{-1}$} (WCL-WCE, respectively).
\end{itemize}
 
\begin{table}
\hspace{1cm}
\begin{minipage}{0.36\textwidth}
\renewcommand{\footnoterule}{}  
\begin{normalsize}
\caption{Average line luminosities for WNL stars}
\label{table:wnl_lum}
\begin{center}
\begin{tabular}{lccc}
\hline \hline
   \noalign{\smallskip}
HeII $\lambda$4686	&	Z (range)	&	Ref.	\\
($\times 10^{35}$ erg s$^{-1}$)	&		&		\\
 \hline
   \noalign{\smallskip}
32	& 	Z$_\odot/3$-Z$_\odot/2$	&	[1]	\\
17	&	Z$_\odot/2$	&	[2]	\\
16 	&	Z$_\odot$	&	[3]	\\
20-26	&	Z $<\mathrm{Z}_\odot$ -- Z  $ \ge \mathrm{Z}_\odot$ 	&	[4]	\\
2-16	&	Z$_\odot/50$ -- Z$_\odot$	&	[5]	\\
4-25	&	Z $< \mathrm{Z}_\odot/5$ -- Z $ \ge \mathrm{Z}_\odot/5$ 	&	[6]	\\
\hline \noalign{\smallskip}
\multicolumn{4}{@{} p{\textwidth} @{}}{ \textbf{Notes.} References: [1]~\cite{Smith91}; [2]~\cite{Vacca92}; [3]~\cite{Schaerer98}; [4]~\cite{Guseva00}; [5]~\cite{Crowther06a}; [6]~\cite{Brinchmann08b}.}
\end{tabular}
\end{center}
\end{normalsize}
\end{minipage}
\end{table}

In \paperone~we estimated the age of the ionizing population in all the identified \hii regions in the disc of NGC 3310. In all cases but two where WR features have been identified, the estimated age for this population is \mbox{$\tau > 3.0$ Myr}. Thus, we do not expect OIf stars to show a significant contribution (if any) to the emission of the broad \heii 4686 \AA{} feature, at least for 16 out of the 18 regions in our sample. We derived the number of WN (N$_{\mathrm{WN}}$) stars assuming that only the WNL type contributes to the luminosity of the broad \heii 4686 \AA{} feature: \mbox{N$_{\mathrm{WN}} \sim$ N$_{\mathrm{WNL}} \sim  L_{\mathrm{obs}} (\textrm{\heii} 4686)$ / $L_{\mathrm{WNL}} (\textrm{\heii} 4686)$}. 

Given the reported metallicity dependence of this broad emission, we have used the approach proposed by~\cite{Lopez-Sanchez10a} to estimate the luminosity of a single WNL ($L_{\mathrm{WNL}} (\textrm{\heii} 4686)$) as a function of metallicity:

\begin{equation}
  L_{\mathrm{WNL}} (\textrm{\heii} 4686) = (-5.430 + 0.812 x) \times 10^{36}~\mathrm{erg s}^{-1}
\end{equation}
with $x$ = 12 + log(O/H). In \paperone~we performed a spectrophotometric analysis to fit the SED of the young ionizing stellar population. Among other characteristics, we could estimate the stellar metallicity of this population. The derived metallicity of this population in \hii regions with identified WR features  is around \mbox{Z $\sim 0.004-0.008$} (\mbox{12 + log(O/H) $\sim$ 8.15--8.45}), thus the derived luminosity of a single WNL star using this prescription is typically \mbox{$L_{\mathrm{WNL}} (\textrm{\heii} 4686) \sim 12-14 \times 10^{35}$ erg s$^{-1}$}. Finally, we have also estimated the effect of non-detected WCE stars on the luminosity of the broad \heii 4686 \AA{} feature. Applying the same criteria we used when fitting the broad components, we could estimate the minimum peak we could be able to detect around the \civ 5808 \AA{} feature (red bump). Thus, assuming this peak and a width of 25 \AA{} (the maximum width allowed in the fit for the broad features, though in the red bump this width 
could be somewhat larger;~\citealt{Schild03}) we could estimate the area under the hypothetical Gaussian representing this feature, that is, the flux of the red bump had we been able to detect it. This allowed us to estimate an upper-limit to the number of WCE stars (a single WCE star contributes with a luminosity at 5808 \AA{} of \mbox{$\sim$ 3.3$\times 10^{36}$ erg s$^{-1}$};~\citealt{Vacca92}), and consequently its contribution to the broad \heii 4686 \AA{} feature. This is typically of the order of 10\%, ranging from 5 to 20\%. Therefore, although our estimation of N$_{\mathrm{WNL}}$ is actually an upper-limit to the real number, in general a maximum systematic offset of only about 10\% to the real number is expected due to the unseen WC population, should it exist. 

Taking all the uncertainties and biases into account, we have made a first order estimation of the number of WNL stars, also identified as the total number of WN stars, in each \hii region with clear WR signatures. As can be seen in Table~\ref{table:wr_numbers}, the \hii regions have typically of the order of a few hundred WN stars (for some of them about half a hundred or even less), being the integrated number above 4000. We can compare the number we have obtained for the so-called Jumbo region (corresponding to the \hii regions ID 1 and 4) with that derived in~\cite{Pastoriza93}. They estimated that around 220 WN4-5 stars could explain the measured luminosity of the broad \mbox{\heii 4686 \AA{}} feature, against \mbox{$\sim$ 670 WN} stars obtained in this study. Actually, given the different calibration they used to estimate the number of WRs and the distance they assumed for NGC 3310 (12.5 Mpc), we should compare the measured flux of the blue bump in both studies. As a matter of fact, the flux we measure 
is larger by a factor of more than 2 
(\mbox{3.4} against \mbox{1.2 $\times 10^{-18}$ erg cm$^{-2}$ s$^{-1}$}). The WR features are normally very weak and can be easily diluted if there is strong presence of non-ionizing stellar emission, as it is the case in the circumnuclear region in NGC 3310. Therefore, a high S/N ratio is required to properly measure the emitted flux in these features. Although the spectral resolution of our spectra is lower than in~\cite{Pastoriza93}, the S/N of the blue bump is much higher with our data in the case of the Jumbo region (compare Fig.~5g in~\citealt{Pastoriza93} and Fig.~\ref{fig:wrb_fits} here).

\begin{table*}
\begin{minipage}{0.93\textwidth}
\renewcommand{\footnoterule}{}  
\begin{normalsize}
\caption{O and WR stellar populations}
\label{table:wr_numbers}
\begin{center}
\begin{tabular}{lcccccccc}
\hline \hline
   \noalign{\smallskip}
HII ID & N$_\mathrm{WNL}$ & $\eta_0$ & N$^\prime _\mathrm{O}$ & N$_\mathrm{O}$ &  $\frac{\mathrm{N}_\mathrm{WNL}}{\mathrm{N}_\mathrm{WNL} + \mathrm{N}^\prime _\mathrm{O}}$ & $\frac{\mathrm{N}_\mathrm{WNL}}{\mathrm{N}_\mathrm{WNL} + \mathrm{N}_\mathrm{O}}$ & $\frac{\mathrm{N}_\mathrm{WR}}{\mathrm{N}_\mathrm{WR} + \mathrm{N}_\mathrm{O}}$ (A89) & $\frac{\mathrm{N}_\mathrm{WNL}}{\mathrm{N}_\mathrm{WNL} + \mathrm{N}_\mathrm{O}}$ (LS10) \\
  (1) & (2) & (3) & (4) & (5) & (6) & (7)  & (8)  & (9)\\
 \hline
   \noalign{\smallskip}
1& 474 $\pm$ 128& 0.45 $^{+0.25}_{-0.23}$ & 3350 $^{+3181}_{-1296}$ & 6622 $^{+6703}_{-2971}$ & 0.118 $^{+0.090}_{-0.064}$ & 0.064 $^{+0.059}_{-0.035}$ & 0.064 $\pm$ 0.030& 0.064 $\pm$ 0.010\\
\noalign{\smallskip}
2& 294 $\pm$ 51& 0.45 $^{+0.45}_{-0.20}$ & 2051 $^{+1173}_{-820}$ & 4499 $^{+2881}_{-1936}$ & 0.125 $^{+0.077}_{-0.047}$ & 0.061 $^{+0.045}_{-0.025}$ & 0.059 $\pm$ 0.020& 0.066 $\pm$ 0.006\\
\noalign{\smallskip}
4& 284 $\pm$ 32& 0.50 $^{+0.30}_{-0.15}$ & 1250 $^{+451}_{-388}$ & 2656 $^{+1227}_{-968}$ & 0.186 $^{+0.071}_{-0.050}$ & 0.096 $^{+0.051}_{-0.030}$ & 0.083 $\pm$ 0.026& 0.071 $\pm$ 0.005\\
\noalign{\smallskip}
5& 382 $\pm$ 76& 0.20 $^{+0.05}_{-0.10}$ & 812 $^{+975}_{-565}$ & 4000 $^{+3379}_{-2310}$ & 0.293 $^{+0.315}_{-0.159}$ & 0.085 $^{+0.107}_{-0.042}$ & 0.143 $\pm$ 0.086& 0.097 $\pm$ 0.010\\
\noalign{\smallskip}
6& 393 $\pm$ 48& 0.21 $^{+0.05}_{-0.05}$ & 434 $^{+486}_{-304}$ & 4138 $^{+1963}_{-1505}$ & 0.455 $^{+0.295}_{-0.193}$ & 0.086 $^{+0.050}_{-0.029}$ & 0.156 $\pm$ 0.058& 0.090 $\pm$ 0.008\\
\noalign{\smallskip}
7& 412 $\pm$ 60& 0.23 $^{+0.05}_{-0.05}$ & 350 $^{+466}_{-252}$ & 2598 $^{+1865}_{-1473}$ & 0.499 $^{+0.293}_{-0.223}$ & 0.136 $^{+0.141}_{-0.057}$ & 0.166 $\pm$ 0.080& 0.096 $\pm$ 0.007\\
\noalign{\smallskip}
8& 428 $\pm$ 37& 0.45 $^{+0.25}_{-0.22}$ & 1165 $^{+1009}_{-430}$ & 3802 $^{+3622}_{-1676}$ & 0.266 $^{+0.110}_{-0.108}$ & 0.101 $^{+0.068}_{-0.047}$ & 0.096 $\pm$ 0.033& 0.072 $\pm$ 0.003\\
\noalign{\smallskip}
9& 209 $\pm$ 101& 0.23 $^{+0.27}_{-0.05}$ & 2326 $^{+1436}_{-1104}$ & 5209 $^{+2577}_{-2114}$ & 0.081 $^{+0.110}_{-0.051}$ & 0.038 $^{+0.041}_{-0.022}$ & 0.072 $\pm$ 0.051& 0.078 $\pm$ 0.024\\
\noalign{\smallskip}
10& 251 $\pm$ 44& 0.45 $^{+0.05}_{-0.05}$ & 176 $^{+190}_{-124}$ & 3427 $^{+526}_{-441}$ & 0.559 $^{+0.266}_{-0.218}$ & 0.068 $^{+0.019}_{-0.016}$ & 0.068 $\pm$ 0.014& 0.061 $\pm$ 0.008\\
\noalign{\smallskip}
11& 328 $\pm$ 66& 0.25 $^{+0.05}_{-0.05}$ & 456 $^{+1110}_{-334}$ & 2135 $^{+4568}_{-1101}$ & 0.366 $^{+0.347}_{-0.240}$ & 0.125 $^{+0.133}_{-0.083}$ & 0.195 $\pm$ 0.098& 0.115 $\pm$ 0.020\\
\noalign{\smallskip}
12& 334 $\pm$ 44& 0.25 $^{+0.05}_{-0.05}$ & 175 $^{+249}_{-129}$ & 1620 $^{+794}_{-661}$ & 0.602 $^{+0.257}_{-0.239}$ & 0.170 $^{+0.106}_{-0.059}$ & 0.235 $\pm$ 0.080& 0.115 $\pm$ 0.013\\
\noalign{\smallskip}
13& 175 $\pm$ 28& 0.70 $^{+0.70}_{-0.35}$ & 291 $^{+248}_{-141}$ & 2574 $^{+1993}_{-1134}$ & 0.372 $^{+0.188}_{-0.147}$ & 0.063 $^{+0.048}_{-0.027}$ & 0.040 $\pm$ 0.008& 0.035 $\pm$ 0.004\\
\noalign{\smallskip}
15& 70 $\pm$ 17& 0.35 $^{+0.30}_{-0.14}$ & 578 $^{+319}_{-228}$ & 1061 $^{+575}_{-416}$ & 0.107 $^{+0.078}_{-0.046}$ & 0.061 $^{+0.045}_{-0.025}$ & 0.070 $\pm$ 0.021& 0.065 $\pm$ 0.009\\
\noalign{\smallskip}
16& 63 $\pm$ 12& 1.20 $^{+0.20}_{-0.55}$ & 180 $^{+71}_{-34}$ & 775 $^{+319}_{-182}$ & 0.252 $^{+0.081}_{-0.080}$ & 0.074 $^{+0.029}_{-0.025}$ & 0.035 $\pm$ 0.005& 0.033 $\pm$ 0.004\\
\noalign{\smallskip}
20& 28 $\pm$ 8& 0.55 $^{+0.55}_{-0.30}$ & 163 $^{+140}_{-73}$ & 496 $^{+523}_{-268}$ & 0.142 $^{+0.121}_{-0.073}$ & 0.052 $^{+0.064}_{-0.028}$ & 0.042 $\pm$ 0.017& 0.029 $\pm$ 0.004\\
\noalign{\smallskip}
21& 51 $\pm$ 9& 0.70 $^{+0.70}_{-0.40}$ & 123 $^{+139}_{-59}$ & 898 $^{+994}_{-417}$ & 0.288 $^{+0.179}_{-0.138}$ & 0.053 $^{+0.046}_{-0.028}$ & 0.037 $\pm$ 0.007& 0.034 $\pm$ 0.004\\
\noalign{\smallskip}
28& 78 $\pm$ 15& 0.25 $^{+0.30}_{-0.05}$ & 94 $^{+120}_{-67}$ & 877 $^{+611}_{-451}$ & 0.415 $^{+0.323}_{-0.207}$ & 0.081 $^{+0.082}_{-0.036}$ & 0.119 $\pm$ 0.073& 0.066 $\pm$ 0.008\\
\noalign{\smallskip}
36& 44 $\pm$ 9& 0.25 $^{+0.05}_{-0.05}$ & 69 $^{+77}_{-48}$ & 536 $^{+281}_{-237}$ & 0.355 $^{+0.321}_{-0.181}$ & 0.075 $^{+0.064}_{-0.031}$ & 0.122 $\pm$ 0.073& 0.104 $\pm$ 0.016\\
\noalign{\smallskip}
\hline \noalign{\smallskip}
\multicolumn{9}{@{} p{\textwidth} @{}}{\textbf{Notes.} Col (1): \hii identification number. Col (2): number of WNL stars. Col (3) adopted $\eta_0$  parameter. Col (4): number of O stars if no absorption by dust grains is taken into account. Col (5): as in Col (4) but a correction due to absorption by dust grains is applied to the \hb luminosity (i.e., to the derived number of O stars. Col (6): number ratio of WNL stars with respect to the total numbers of O and WNL stars if no absorption by dust grains is taken into account. Col (7): as in Col (6) but a correction due to absorption by dust grains is applied to the \hb luminosity. Col (8): same ratio for WRs using the calibration proposed by~\cite{Arnault89} (A89). Col (9): same ratio for WNL stars but using the calibration proposed by~\cite{Lopez-Sanchez10a} (LS10).}
\end{tabular}
\end{center}
\end{normalsize}
\end{minipage}
\end{table*}

\subsection{WR ratios}
\label{sec:wr_ratios}

Empirical results such as the ratio of WR to O stars provide sensitive tests of evolutionary models which involve complex processes (i.e., rotation, binarity). This ratio can be roughly derived by first estimating the number of O stars using the \hb luminosity. Most of the \hii regions in NGC 3310 have low \oiii/\oii~ratios (\mbox{i.e., $< 1$};~\paperone). According to~\cite{Osterbrock82}, if the observed level of ionization is low, O7V stars are the best representative type of O stars. Assuming a contribution to \hb luminosity by an O7V star of \mbox{$L_{\mathrm{O7V}} = 4.76\times10^{36}$ erg s$^{-1}$}, a first estimation of the number of such stars is directly \mbox{N$_{\mathrm{O7V}} = L(\mathrm{H}\beta)/L_{\mathrm{O7V}}$}.  
However, the contribution of the WR and other O stars subtypes to the ionizing flux has to be taken into account:

\begin{itemize}
 \item Following~\cite{Crowther06a}, the average number of ionizing photons of a WNL star is assumed to be log $Q_{0}^{\mathrm{WNL}} = 49.4$.   
 \item The total number of O stars (N$_\mathrm{O}$) can be derived from the number of O7V (N$_\mathrm{O7V}$) stars by correcting for other O stars subtypes, using the parameter $\eta_0$ introduced by~\cite{Vacca92} and~\cite{Vacca94}. This parameter depends on the initial mass function for massive stars and is a function of time because of their secular evolution~\citep{Schaerer98}. With our estimation of the age of the ionizing population for each \hii region in NGC 3310 as reported in \mbox{Paper~I}, we estimated $\eta_0$ using the SV98 models for a metallicity \mbox{Z = 0.4Z$_\odot$} (the typical metallicity observed in this galaxy;~\citealt{Pastoriza93}, \mbox{Paper~I}). The strongly non-linear temporal evolution of this parameter during some time intervals (see Fig.~21 in~\citealt{Schaerer98}) causes strong asymmetries in the determination of its uncertainty (see Table~\ref{table:wr_numbers}).  
\end{itemize}

\begin{figure*}
\centering
\includegraphics[angle=90,trim = 3.5cm 0cm 4cm -1cm,clip=true,width=0.9\textwidth]{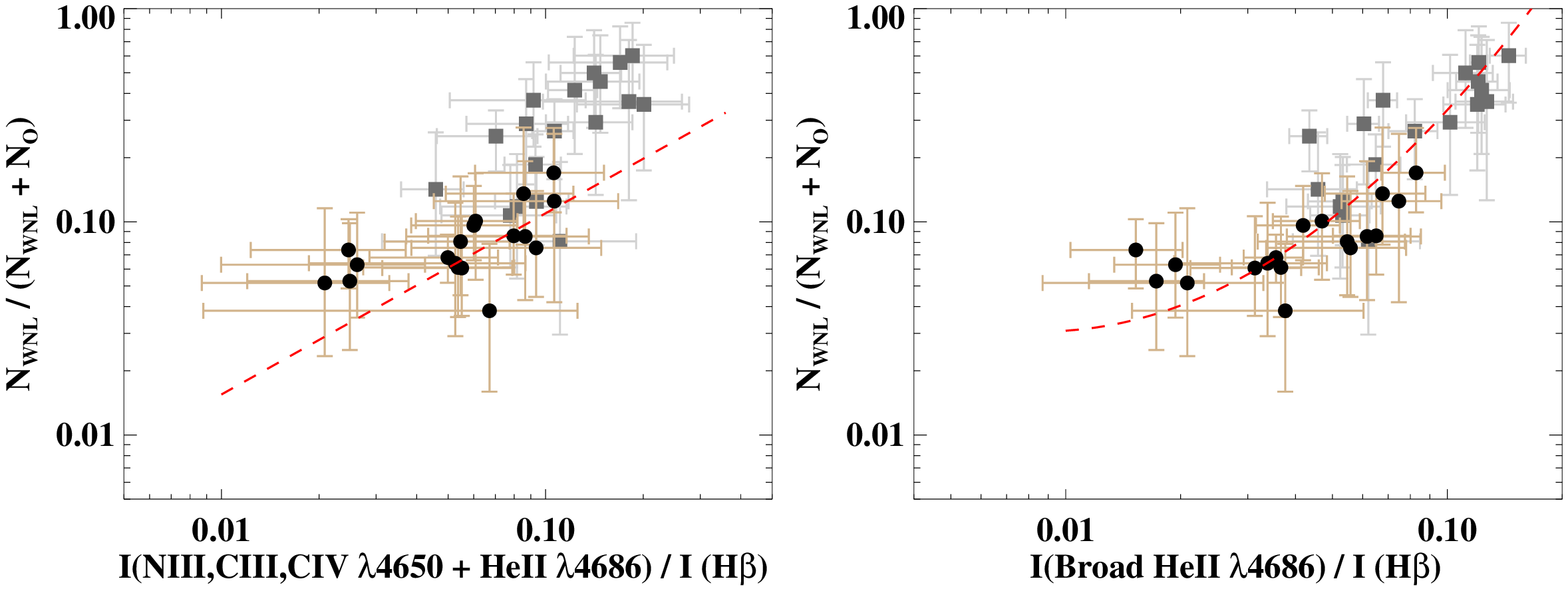}
  \caption{\textbf{Left:} Number of WNL to WNL+O ratio (as defined in the text) vs. the reddening corrected flux of the broad \mbox{\heii 4686 \AA{}} line + blend centred at 4650\AA{} in \hb units. The empirical relation given by Schaerer \& Vacca (\citeyear{Schaerer98}) is drawn in a dashed red line. \textbf{Right:} The same number ratio vs. the reddening corrected flux of just the broad \mbox{\heii 4686 \AA{}} line to \hb. The empirical relation obtained by L\'opez-S\'anchez \& Esteban~(\citeyear{Lopez-Sanchez10a}) is shown with a dashed red line. In both cases, squares in grey refer to data not corrected by absorption by dust grains in the nebula ($f_d$ in \mbox{Paper I}), while dots in black refer to data corrected by this absorption.}
  \label{fig:wr_ratios}
\end{figure*}

With this we can determine the number of O stars as:

\begin{equation}
 \mathrm{N}_\mathrm{O} = \frac{Q_0^{\mathrm{Total}} - N_\mathrm{WNL}Q_0^{\mathrm{WNL}}}{\eta_0 Q_0^{\mathrm{O7V}}}
\end{equation}
where $Q_0^{\mathrm{Total}} = \mathrm{N}_{\mathrm{O7V}} Q_0^{\mathrm{O7V}} $ and $Q_0^{\mathrm{O7V}}$ are the total and O7V number of ionizing photons, respectively. We have adopted an average Lyman continuum flux per O7V star of \mbox{log $Q_0^{\mathrm{O7V}} = 49.0$} (\citealt{Vacca92,Schaerer98,Schaerer99}). As Table~\ref{table:wr_numbers} shows, the number of WNL stars spans over one order of magnitude, depending on the \hii regions, from about 30 to 500. In \paperone~ we performed a spectro-photometric and ionization model analysis of the identified \hii regions in NGC 3310 and, as a result, it was determined that at least 25\% of the ionizing photons are absorbed by dust in the nebula (with dust absorption factors $f_d \gtrsim 1.5$ for regions with identified WR features). Therefore, we have derived the number of O stars applying a correction factor to the derived Q(H) due to this effect (N$_\mathrm{O}$) and without the correction (N$_\mathrm{O}'$). As can be seen in Table~\ref{table:wr_numbers}, 
differences in the derived number of O stars taking into account UV dust absorption or not range from about a factor of 2 up to an order of magnitude. Therefore, this absorption can have a major impact on the determination of the number of ionizing stars and, as a consequence on the WR/O ratio. We must also keep in mind that due to the faintness of the broad \civ 5808 \AA{} lines, some WC stars contribution may be also expected, and hence the total number of WR (O) stars may well represent just a lower (upper) limit to the actual value. If we assumed that there are a maximum of about 10\% of unseen WC stars, then for a ratio \mbox{$\frac{\mathrm{N}_\mathrm{WNL}}{\mathrm{N}_\mathrm{WNL} + \mathrm{N}_\mathrm{O}}$ = 0.10} the corrected value would be around 0.08--0.09 depending on $\eta_0$. For lower ratios, the impact would even be less important.

\begin{figure*}
\centering
\includegraphics[angle=90,trim = 4cm 0cm 4cm -1cm,clip=true,width=0.95\textwidth]{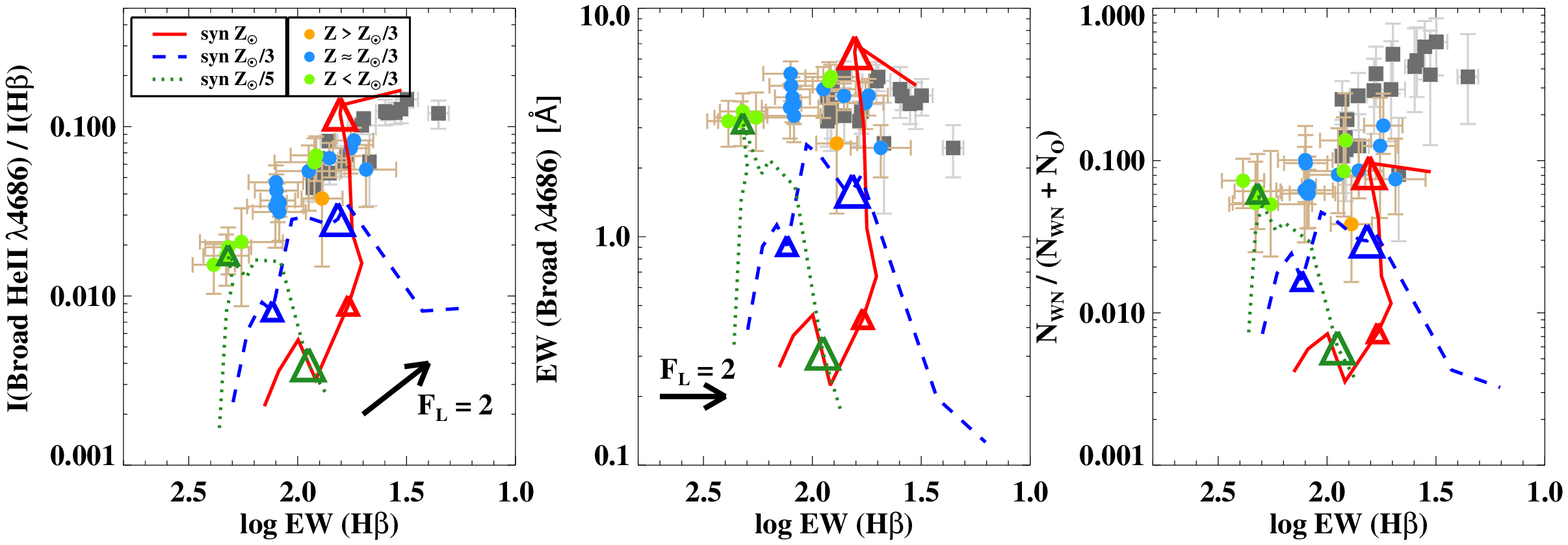}
  \caption{HeII~ $\lambda 4686$ intensity (left), EW (centre) and the derived ratio of number of WRs over the total number of massive WR + O stars (right) vs. EW (\hb\onespace). Filled circles refer to the data corrected by the emission of a non-ionizing underlying population and absorption by dust grains within the nebula. They are differently coloured depending on the derived metallicity in \mbox{Paper~I} (see legend on the left-hand side plot). Gray squares refer to the data with no correction by the mentioned dust absorption (as in Fig.~\ref{fig:wr_ratios}). Tracks from \popstar have been overplotted in different colours and type of line (see legend) depending on metallicity. Open triangles mark the values for 4 (small triangles) and 5 Myr (large triangles), respectively, on the tracks. The arrows labelled as \mbox{F$_\mathrm{L}$ = 2} illustrate how the tracks would move on the plot if half of the ionizing photons actually escape the ionized regions.}
  \label{fig:wr_vargas}
\end{figure*}

We have compared our WR/O ratio, as the number of WRs over the total number WR+O stars (N$_\mathrm{WR}/($N$_\mathrm{WR}$+N$_\mathrm{O}$)), with a couple of calibrations using either the  4650 blend (i.e., the entire WR blue bump) or just the \heii 4686 \AA{} broad line. The former was early proposed by~\cite{Arnault89} and is useful especially in low resolution spectra, for which the broad \heii 4686\AA{} cannot be separated clearly from the bump. The calibration is:

\begin{equation}
 \mathrm{log}\left(\frac{\mathrm{N}_\mathrm{WR}}{\mathrm{N}_\mathrm{WR} + \mathrm{N}_\mathrm{O}} \right) = (-0.11 \pm 0.02) + (0.85 \pm 0.02) \mathrm{log} \left(\frac{L_{\mathrm{4650}}}{L_{\mathrm{H}\beta}} \right)
\end{equation}
 
On the other hand, using data from~\cite{Crowther06a},~\cite{Lopez-Sanchez10a} developed an empirical calibration  of the WR/O ratio using only the intensity of the broad \heii 4686 \AA{} line, for the case of the WNL subtype:

\begin{equation}
 \mathrm{log}\left(\frac{\mathrm{N}_\mathrm{WNL}}{\mathrm{N}_\mathrm{WNL} + \mathrm{N}_\mathrm{O}} \right) = -1.511 + 0.1162 x + 0.9194 x^2
\end{equation}
where \mbox{$x$ = log [$I$(\heii 4686)/$I$(H$\beta$)]}, normalizing to \mbox{$I$(H$\beta$) = 100}.

We compare our results with both calibrations in Fig.~\ref{fig:wr_ratios}. If absorption by dust grains is not taken into account (grey squares in the figure), most of the values lie away from the relation given by~\cite{Arnault89} and the ratios obtained are compatible with having more WR than O-type stars in some cases. This also differs dramatically from observations, given that the largest ratios observed in WR galaxies are of the order or 0.10-0.30~\citep{Castellanos02,Crowther06a,Brinchmann08a,Perez-Montero10,Lopez-Sanchez10a}. If, on the other hand, absorption is allowed, then the corrected values (black points in figure) are more consistent with the empirical relations and observations, though we observe a rather constant ratio for intensity ratios (for both \mbox{$I$(\nii, \ciii, \civ 4650 + \heii 4686)/$I$(H$\beta$)} and \mbox{$I$(\heii 4686)/$I$(H$\beta$)}) lower \mbox{than $\sim 0.06$}. For a few of the \hii regions there may still be a systematic offset of about 0.1 dex between our derived 
values and the calibration proposed by~\cite{Arnault89}. This is actually expected since, as 
we have mentioned before, the total number of WRs can well be somewhat underestimated and the number of O stars overestimated. In fact, the 4650 feature can include emission from carbon (\ciii and \civ) lines, originated in WC stars.

\section[]{Discussion}
\label{sec:discussion}

\subsection{Comparison with stellar population models}

During the last 20 years theoretical models have tried to explain the observed emission of the WR population, such as the measurements derived in the previous section. Starting with~\cite{Arnault89}, and up to now, improvements have been achieved by the combination of stellar evolution models, theoretical spectra and compilations of observed line-strengths from WRs, the most widely used those by~\cite{Schaerer98} (SV98). Nowadays, there is a better understanding of the properties of WRs than when the SV98 models were released. Some improvements have been achieved on the metallicity-dependence of the \heii 4686 luminosity~\citep{Crowther06a}, on the influence of the rotation~\citep{Meynet05,Leitherer14} and on the effect of wind loss and binary evolution~\citep{vanBever03,vanBever07,Eldridge08,Eldridge09} in the stellar tracks followed by massive stars.  

Here we compare the predictions of theoretical models with our observational data. In particular, two models are considered for the comparison: (i) \popstar~\citep{Molla09,Martin-Manjon10}, a self-consistent set of models including the chemical and the spectro-photometric evolution, for spiral and irregular galaxies, where star formation and dust effects are important; and (ii) \bpass~\citep{Eldridge09}, which includes the binary evolution in modelling the stellar populations, that can extend the WR phase up to longer than 10 Myr. Both models use the photoionization code  \cloudy~\citep{Ferland98} to predict the nebular emission.

\begin{figure*}
\centering
\includegraphics[angle=90,trim = 4cm 0cm 4cm -1cm,clip=true,width=0.95\textwidth]{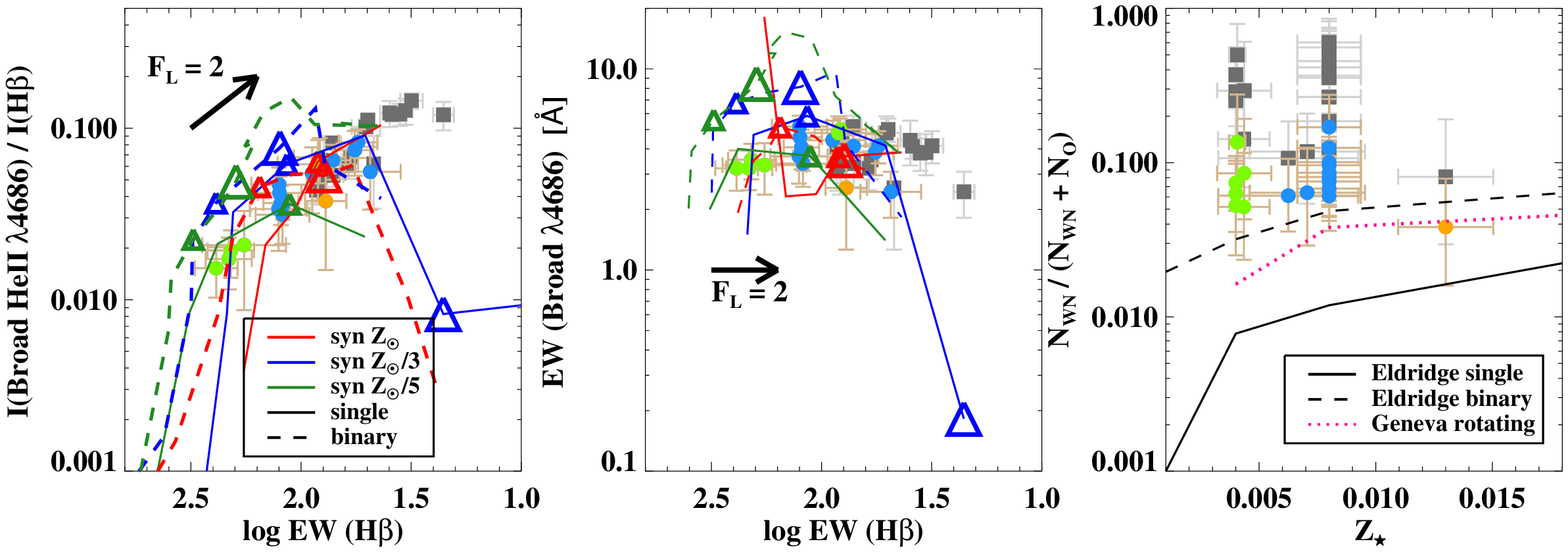}
  \caption{\textbf{Left \& centre:} HeII~ $\lambda 4686$ intensity (left), EW (centre) vs. EW (\hb\onespace), as in Fig.~\ref{fig:wr_vargas}. Now, the BPASS models are overplotted, which include binarity; the colours and type of line for the models differ from those in Fig.~\ref{fig:wr_vargas}, as the legend on the left-hand side plot specifies. This time the open triangles mark the values for 5 (small triangles) and 8 (large triangles) Myr on the tracks. \textbf{Right:} Derived ratio of number of WRs over the total number of massive WR + O stars (right) vs. derived \textit{stellar} metallicity in \paperone. Tracks from BPASS models are also overplotted. Additionally, the dotted pink line in the right plot shows the Geneva track for rotating stars from Meynet \& Maeder (\citeyear{Meynet05}).}
  \label{fig:wr_eldridge}
\end{figure*}

We first compare the intensity ratio of the blue bump to the \hb emission line, the blue bump EW and the derived WR to O number ratio (i.e., N$_{\mathrm{WN}}/(\mathrm{N}_{\mathrm{WN}} + \mathrm{N}_{\mathrm{O}}$) vs. the EW (\hb\onespace) (i.e., basically the age of the population), with predictions from \popstar models (Fig.~\ref{fig:wr_vargas}). These models were used in \mbox{Paper~I} to estimate the age of the ionizing population. As can be seen in the figure, for the typical moderate metallicities observed in this galaxy (i.e., \mbox{$Z \lesssim Z_\odot/3$}), evolutionary tracks from models lie in general well below the observed values, by factors larger than 3 (grey squares in the figure). If we correct the observed values by dust absorption of UV photons in the nebula (coloured dots in the figure), models and observations get closer, though a typical mismatch of factors of 2--3 is still evident\mbox{, specially at moderate metallicities (blue dots)}. We have also included in the plots vectors 
indicating how the tracks would move if half the 
ionizing photons escaped from the \hii regions. This is motivated by the results of some works in extragalactic \hii regions which conclude that the Lyman continuum photon leakage fraction may be significant (e.g.,\citealt{Castellanos02b,Iglesias-Paramo02,Giammanco05,Grossi10}). Although the correction due to Lyman photon leakage to the model tracks would imply that the tracks would have larger WR to O number ratios, we cannot quatify the effect since the number of O stars depends on the derived number of WRs. If escape of ionizing photons occurred, then the modelled tracks would move in the same direction as if correcting them by the absorption of UV photons, and thus the situation would not change perceptibly for the first two plots.

The mismatches between models and data have been known for some time now (e.g.,~\citealt{Guseva00,Crowther06a,Perez-Montero10,Lopez-Sanchez10a}), especially when trying to explain the derived large WR to O ratios from observations (although in order to derive the number of O stars model prescriptions are used; see Sect.~\ref{sec:wr_ratios}). As discussed in~\cite{Crowther07}, the production of more WRs than currently favoured by the models can be achieved by including binarity in the evolutionary codes or when rotation is included in the stellar tracks. They have become promising sources for an increased WR population. In fact, several studies on samples of Galactic massive O stars  support that binary interaction dominates the evolution of massive stars (e.g.~\citealt{Kobulnicky07,Sana12,Kiminki12}).

\cite{Eldridge08} developed a synthesis population code, BPASS\footnote{http://www.bpass.org.uk/}, where binarity is included. They found out that a third of the population evolves as single stars, while the remaining two thirds correspond to interacting binaries. The inclusion of binaries led to a prolonged WR phase (up to \mbox{$\tau \sim$ 15 Myr}), consistent with earlier predictions by~\cite{vanBever03}.  Fig.~\ref{fig:wr_eldridge} compares our measurements and estimates with those predicted by BPASS, for single and binary star models. These models do not provide the WR ratio as a function of the age of the burst (i.e.,\mbox{~EW (\hb\onespace))}. Nevertheless, they explicitly show that taking into account binaries leads to a increase in the WR to O number ratio relative to single stars models (Fig.~\ref{fig:wr_eldridge}, right). This had been a long-standing problem:  given that metallicity  has a strong effect on mass-loss rates, as WRs typically need a period of high mass loss to form, very few 
of them were expected to form at metallicities much lower than solar; however many WRs have been observed at low metallicites (e.g.,~\citealt{Schaerer98,Royer01,Brinchmann08a}).

\begin{figure*}
\centering
\includegraphics[angle=90,trim = 2cm 0cm 3cm -1cm,clip=true,width=0.95\textwidth]{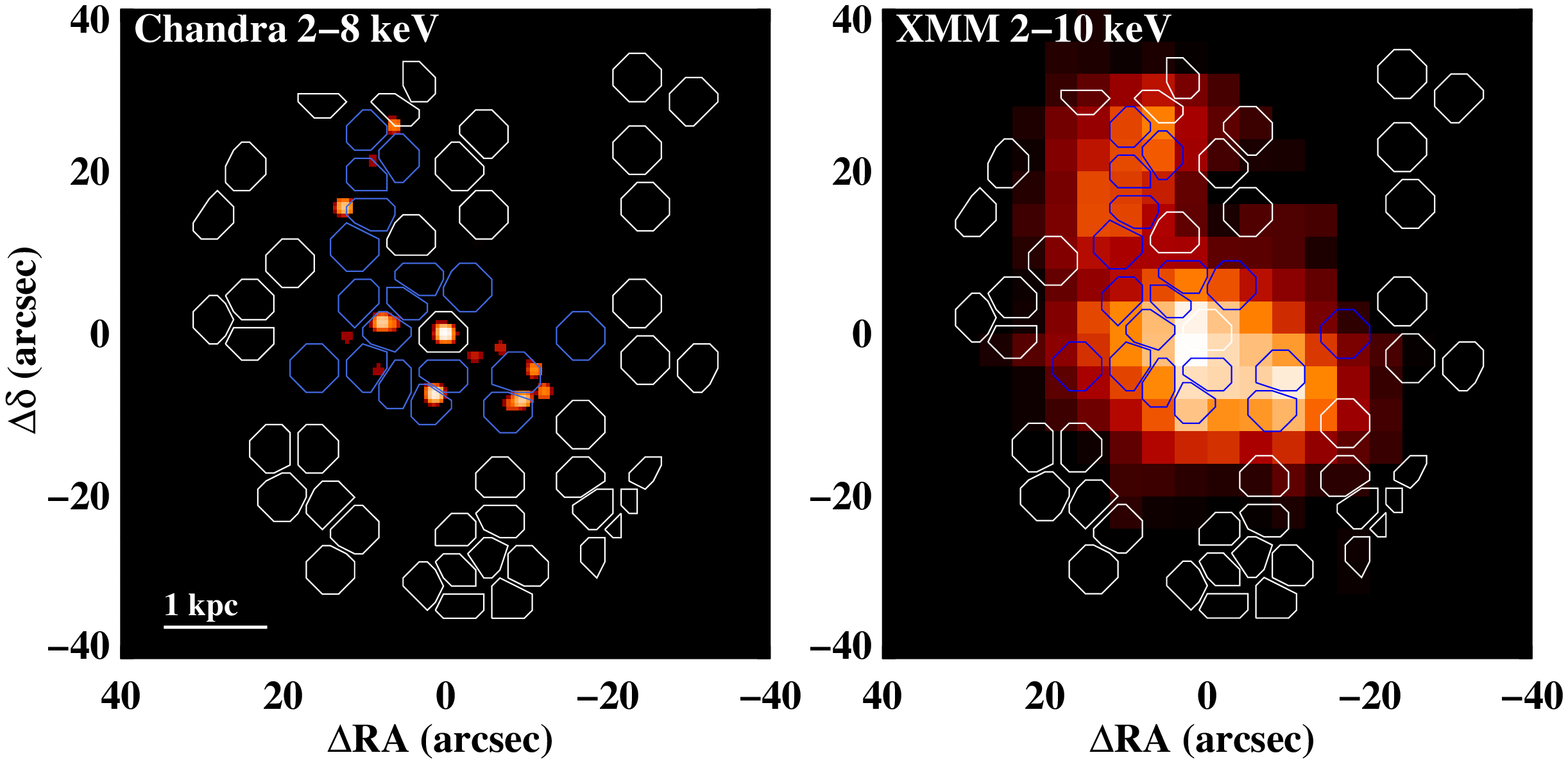}
  \caption{\textbf{Left:} \textit{Chandra} X-ray image in the 2--8 keV energy range of the central region of NGC 3310 where \hii regions with measurable WR features have been found. The aperture that defines each \hii region as shown in Paper I is overplotted in blue for \hii regions with detected blue bump and in white for the rest. The horizontal line at the bottom left corner shows a scale of 1 kpc. \textbf{Right:} \textit{XMM-Newton} X-ray image in the 2--10 keV energy range of the same region. The aperture that defines each \hii region is also overplotted.}
  \label{fig:x_ray_im}
\end{figure*}

As before, if absorption by dust grain is not taken into account, observations and predictions differ in general. Once a correction for this is applied to the observed data, we can see that:
\begin{itemize}
 \item The observed values (intensity ratio and blue bump equivalent width) are in general consistent with single star models, having similar ages as derived in \mbox{Paper~I}. However, predicted WR to O number ratios are significantly underestimated, by up to factors of ten or larger, \text{if single star models are consdidered (Fig.~\ref{fig:wr_eldridge}, right)}.  

 \item Including the effect of active binaries in the models enhances the intensity ratio, the EW and the WR to O ratio. In some cases the observed values are consistent with binary star models, the implied ages being typically of the order of 10 Myr or older. In other cases, corrections (e.g., due to escape of ionizing photons) should be applied to the tracks in order to match observations and binary star models. In such cases, the derived ages would be more consistent with those estimated in \mbox{Paper~I}. Regarding the WR to O number ratios, our derived values and predictions using the BPASS models are consistent in some cases while in others they differ by factors of 2--4. 

 \item The effect of fast rotation is also showed in the figure (right). A combination of rotation and binarity may even enhance more the number of predicted WRs. 
\end{itemize}

Finally, aperture effects may also affect the measurements. As discussed in~\cite{Kehrig13}, WRs are normally localized in a small area compared to the more extended presence of the rest of the ionizing population. If a too-small aperture is used all the light from WRs is measured but not all the light coming from the rest of the ionizing population. On the other hand, if a too-large aperture is applied, dilution of the light from WRs occurs because the continuum increases and light from other ionizing populations may be included. The former happened with the spectrum from SDSS obtained for the metal-poor WR galaxy \mbox{Mrk 178} (\citealt{Kehrig13}). This study could be affected by the latter, given that the typical size of the \hii regions in this study is of the order of 150--250 pc, typical size of giant \hii regions, where several ionizing populations may be present (e.g., like in 30 Doradus;~\citealt{Doran13}). Should this happen, ratios of properties of the WR population to those of the 
whole ionizing population (i.e., WR to O number ratio, intensity ratio of the blue bump flux at $\lambda$4686 to \hb, equivalent width of the blue bump) would be diminished due to aperture effects, which would result in even worse agreement with models. 

All in all, it is clear that a single model is not able to reproduce the observed values in many cases. Nevertheless, if additional processes are taken into account (i.e., dust absorption, photon leakage, binarity, fast rotation of the WRs), a general agreement could be reached.

\subsection{WRs and binarity}

In the previous section we compared our observations with different stellar population models. Models that include binary interactions and/or rotation are able to better reproduce some observables. Therefore, motivated by the models, in this section we investigate the binary channel by studying the X-ray emission in the circumnuclear region of NGC 3310. That way we can test the binary fraction of the ionizing population with strong presence of WRs.

The X-ray emission of starburst galaxies is mainly produced by high-mass X-ray binaries (HMXB), supernova remnants (SNR) and hot gas heated by both the energy originated in supernova (SN) explosions and stellar winds~\citep{Cervinyo02,Persic02,Fabbiano06}. While most of the soft part of the X-ray emission (\mbox{0.5--2 keV}) is produced by gas at \mbox{kT $\sim$ 0.3-0.7 keV}, in general the hard X-ray emission (\mbox{2--10 keV}) is dominated by High Mass X-ray Binaries, HMXB,~\citep{Persic04}. They are compact objects which are accreting matter from a massive companion star. When the  matter of the disc falls onto the compact object, the potential energy released ends up heating the gas up to millions of Kelvin, generating hard X-ray emission. Active Galactic Nuclei (AGNs) are also conspicuous hard X-ray emitters, and can dominate the emission of their host galaxy, and indeed, hard X-ray emission has been used as a tracer of AGN activity in galaxies. If there is no AGN, the hard X-ray luminosity from HMXBs 
can be used as a 
star formation tracer~\citep{Grimm03,Ranalli03,Persic04,Lehmer10,Pereira-Santaella11,Mineo12}.

Fig.~\ref{fig:x_ray_im} shows a 0.5\arcsec~ resolution X-ray image from the Chandra X-ray Observatory in the \mbox{2--8 keV} energy range (left), and a lower resolution XMM-Newton X-ray image in the  (\mbox{2--10 keV}) energy range. Different components can be observed in the former image while only about three blended knots of X-ray emission are distinguished in the latter. \cite{Tzanavaris07} obtained the spectrum of the nucleus of NGC 3310 using Chandra data. They could fit their spectrum with an empirical power-law, a {\small MEKAL} model (of a soft thermal or hot plasma) and moderate intrinsic absorption. The thermal plasma represents the soft X-ray emitting gas heated by SN shocks and stellar winds, whereas the power-law reproduces the observed hard X-ray continuum produced by X-ray binaries and/or an AGN. The integration of a Gaussian corresponding to the \mbox{Fe K$\alpha$} emission was discussed there. In fact, although the \mbox{Fe K$\alpha$} emission line has been found in a few starbursts like 
M82 and NGC 253\mbox{~\citep{Cappi99}}, being its origin associated with X-ray binaries and SNRs, this high-energy line is detected mainly in active galaxies. Tzanavaris and their colleagues derived an X-ray unabsorbed luminosity of the central source of about \mbox{2.3$\times 10^{40}$ erg s$^{-1}$} in the range \mbox{0.2--10 keV}, with an uncertainty of a factor of 2. On the other hand, ~\cite{Jenkins04} used XMM data to fit the spectra for the main body of the galaxy (bassically covering an aperture with similar area to that covered in Fig.~\ref{fig:x_ray_im}), and determined an unabsorbed luminosity of about \mbox{1.26$\times 10^{41}$ erg s$^{-1}$} in the range \mbox{0.3--10 keV}. The X-ray flux of the central source is therefore between 10 and 20\% of the total galactic flux in the 0.2--10 keV band, as claimed in~\cite{Tzanavaris07}.

\begin{table*}
\begin{minipage}{\textwidth}
\renewcommand{\footnoterule}{}  
\begin{small}
\caption{Spectral fitting results for NGC 3310.}
\label{table:x_ray_fit}
\begin{center}
\begin{tabular}{@{\hspace{0.16cm}}l@{\hspace{0.16cm}}c@{\hspace{0.16cm}}@{\hspace{0.16cm}}c@{\hspace{0.16cm}}@{\hspace{0.16cm}}c@{\hspace{0.16cm}}@{\hspace{0.16cm}}c@{\hspace{0.16cm}}@{\hspace{0.16cm}}c@{\hspace{0.16cm}}@{\hspace{0.16cm}}c@{\hspace{0.16cm}}@{\hspace{0.16cm}}c@{\hspace{0.16cm}}@{\hspace{0.16cm}}c@{\hspace{0.16cm}}@{\hspace{0.16cm}}c@{\hspace{0.16cm}}@{\hspace{0.16cm}}}
\hline \hline
   \noalign{\smallskip}
Obs ID	&	Fit	&	N$_{\mathrm{H}}$ (PL)	& 	$\Gamma$	&	N$_{\mathrm{H}}$ ({\small VMEKAL})	&	KT	&	[Fe/O]	&	$\chi^2$/d.o.f.	&	Flux (2--10 keV)	&	$L_{\mathrm{X}}$	\\
	&		&	 ( 10$^{21}$ cm$^{-2}$)	& 		&	 ( 10$^{21}$ cm$^{-2}$)	&		&		&		&	 (10$^{-13}$ cgs)	&	(10$^{40}$ erg s$^{-1}$)	\\
(1)	&	(2)	&	(3)	& 	(4)	&	(5)	&	(6)	&	(7)	&	(8)	&	(9)	&	(10)	\\
 \hline
   \noalign{\smallskip}
0556280101	&	Simultaneous	&	0.8	& 	1.65 $\pm$ 0.04	&	1.2 $\pm$ 0.7	&	0.37 $\pm$ 0.05	&	0.51 $\pm$ 0.12	&	1.15	&	9.8 $\pm$ 0.5	&	3.0 $\pm$ 0.2	\\
0556280101	&	Combined	&	1.8	& 	1.74 $\pm$ 0.05	&	\ldots	&	0.42 $\pm$ 0.03	&	0.38 $\pm$ 0.06	&	1.45	&	10.5 $\pm$ 0.4	&	3.3 $\pm$ 0.1	\\
	&		&		& 		&		&		&		&		&		&		\\
0556280102	&	Simultaneous	&	1.0	& 	1.63 $\pm$ 0.05	&	0.1 $^{+0.2}_{-0.1}$ 	&	0.51 $\pm$ 0.03	&	0.37 $\pm$ 0.05	&	1.12	&	9.5 $\pm$ 0.6	&	2.9 $\pm$ 0.2	\\
0556280102	&	Combined	&	1.0	& 	1.64 $\pm$ 0.05	&	\ldots	&	0.51 $\pm$ 0.02	&	0.38 $\pm$ 0.02	&	1.36	&	9.3 $\pm$ 0.3	&	2.9 $\pm$ 0.1	\\
\hline \noalign{\smallskip}
\multicolumn{10}{@{} p{\textwidth} @{}}{\textbf{Notes.} Col (1): ID of the observation. Col (2): fitted spectrum, either a simultaneous fit of the MOS and pn spectra (``simultaneous'') or a fit of the combined spectrum (``combined''). Col (3): column density of the absorption for the power-law component. Col (4): the power-law slope. Col (5): column density of the absorption for the hot plasma component. Col (6): temperature of the thermal component. Col (7): abundance ratio of the thermal component. Col (8): reduced $\chi^2$. Col (9): unabsorbed X-ray flux in the energy range 2--10 keV. The modelled flux for the pn data (the pn camera has higher sensitivity) has been adopted. Col (10): unabsorbed X-ray luminosity in the energy range 2--10 keV, assuming a distance of 16.1 Mpc for the galaxy.}
\end{tabular}
\end{center}
\end{small}
\end{minipage}
\end{table*}

About half of the \hii regions with detectable WR features have an X-ray counterpart (see Fig.~\ref{fig:x_ray_im}, left). We have obtained the EPIC X-ray spectra taken with the two MOS and the pn camera within an aperture of 35\arcsec~in radius, large enough to include all the X-ray emission showed in the figure. Since we have the data corresponding to two observation runs we have total of six spectra to analyse. We fitted each set of three spectra simultaneously using the XSPEC package\footnote{http://heasarc.nasa.gov/xanadu/xspec/}, version 12.8.~\cite{Jenkins04} produced several fits to the spectrum of this source by first fitting it with an absorbed powerlaw, then fixing the column density and finally adding one or two hot plasma (i.e., {\small MEKAL}) components. That way, they ensured a more physically realistic fit (i.e., point sources in star-forming galaxies typically have non-negligible absorption;~\citealt{Lira02}). Following this procedure we successfully fitted the spectra using a 
simple model consisting of an absorbed ({
\small WABS}) thermal plasma ({\small VMEKAL}) plus an absorbed power-law. Contrary to the previously mentioned hot plasma model, {\small VMEKAL} allows to set the individual abundances. The [Fe/O] ratio was left as a free parameter, since the most prominent spectral features in the soft X-ray range are produced by these elements (the Fe L-shell and the O K-shell), and the abundance ratio could be determined. For each set of spectra the parameters where bound, except for the normalization constants, since flux measurements with the MOS cameras are somewhat larger than those made with the pn camera due to calibration issues~\citep{Stuhlinger06}. The results of the fits in a spectral range of \mbox{0.5--8 keV} for both datasets are reported in table~\ref{table:x_ray_fit} (simultaneous fit). It can be seen that, although the absorption of the thermal component is not very well constrained, the slope of the powerlaw is consistent in both fits. The derived slopes and X-ray luminosities are also 
consistent 
with each other. Given that the normalization constants are not the same we could made different determinations of the X-ray flux for each of the three cameras. However, we adopted the flux for the data taken with the pn camera because of its superior sensitivity.  

\begin{figure}
\centering
\includegraphics[angle=-90,trim = 0cm 0.2cm 0.2cm 1.2cm,clip=true,width=1.0\columnwidth]{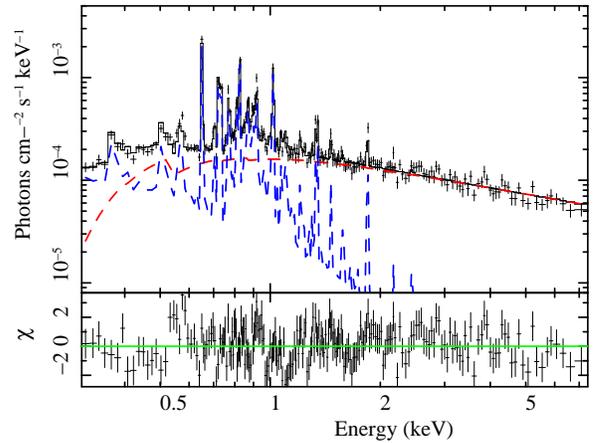}
  \caption{EPIC X-ray spectrum (black crosses) of the circumnuclear region in NGC 3310. Total fitted spectrum is drawn in solid black line, while the components of the spectrum are drawn in dashed blue ({\small VMEKAL}) and red (power-law) lines. Residuals are shown below.}
  \label{fig:x_ray_spe}
\end{figure}

We also produced for each dataset a single spectrum by merging the spectra of the three cameras using the {\small SAS} task \textit{epicspeccombine}. The results of the fits are also shown in the table. We also show in Fig.~\ref{fig:x_ray_spe} an example of the fit for the dataset ID  0556280102. In this case, an absorption to the thermal component could not be constrained and actually dit not improve the fit. Additionally, the fits are a bit poorer, probably due to the systematics between the MOS and the pn cameras. 

The derived power-law component in all cases is slightly steeper than in local starbursts ($\Gamma = 1.2$;~\citealt{Persic03}), and more similar to that in local Luminous Infrared Galaxies, LIRGs ($\Gamma \sim 1.8$). The measured $N_H$ corresponds to $A_V \sim 0.5$ mag using the~\cite{Guver09} conversion factor. This absorption is consistent with that derived at optical wavelengths ($A_V \sim 0.4-0.7$; \mbox{Paper~I}). We adopt an average X-ray luminosity \mbox{$L_{2-10~\mathrm{keV}} = 3 \times 10^{40}$ erg s$^{-1}$}. As mentioned before, \cite{Tzanavaris07} estimated that the contribution to the total X-ray emission of the circumnuclear region in this galaxy of the nucleus is about 10\%. Since we are interested in the high energy regime we compared the X-ray emission of the central source (reported by~\citealt{Tzanavaris07}) with that for the circumnuclear region we measure, but for the power-law component. They obtained \mbox{$L_{\mathrm{X}} (PL; 0.2 - 10 \mathrm{keV}) = 1.95^{+1.69}_{-0.63} \times 10^{40}
$ erg s$^{-1}$}, which at their assumed distance for NGC 3310 gives a flux of about \mbox{$4.7 \times 10^{-13}$ erg cm$^{-2}$ s$^{-1}$}. We derive a flux from \mbox{16 to 20 $\times 10^{-13}$ erg cm$^{-2}$ s$^{-1}$}, about a factor of 4 higher. Therefore, in the high energy regime we estimate the contribution of the central source as about 25\% of the total emission of the circumnuclear region of NGC 3310. Under this assumption, the non-nuclear hard X-ray luminosity (mainly associated with ionizing populations with WR features) corresponds to about $L_{2-10~\mathrm{keV}} = 2.2\times 10^{40}$ erg s$^{-1}$.  

We have applied the well-known correlation between the hard X-ray luminosity and the SFR in order to estimate the total SFR over the last 100 Myr of those regions with an X-ray counterpart, using the prescription by~\cite{Grimm03} for $L_{\mathrm{X}}  \lesssim 3\times 10^{40}$ erg s$^{-1}$:

\begin{equation}
 \mathrm{SFR}~(M_\odot \mathrm{yr}^{-1}) = \left( \frac{L_{2-10~\mathrm{keV}}}{2.6\times10^{39}~\mathrm{erg s}^{-1}}\right)^{0.6} 
\end{equation}
 
Then, \mbox{SFR $ \sim 3.6~M_\odot \mathrm{yr}^{-1}$}. There has recently been proposed another calibration between the SFR and the X-ray luminosity by~\cite{Mineo12}:

\begin{equation}
 \mathrm{SFR}~(M_\odot \mathrm{yr}^{-1}) = \frac{L_{0.5-8~\mathrm{keV}}}{2.61\times10^{39}~\mathrm{erg s}^{-1}} 
\end{equation}

With a flux of \mbox{$F_{\mathrm{X}}~(0.5-8~\mathrm{keV}) \sim 1.3 \times 10^{-12}$ erg cm$^{-2}$ s$^{-1}$}, applying the same correction for the central source of the galaxy, we obtain $L_{0.5-8~\mathrm{keV}} \sim 3\times 10^{40}$ erg s$^{-1}$ and SFR $ \sim 11.5~M_\odot \mathrm{yr}^{-1}$. Using optical data in \paperone~we derived an integrated SFR between 4.4 and \mbox{13 $M_\odot \mathrm{yr}^{-1}$}. Our estimated values using X-ray data lie within the same interval, which is not surprising taking into account that HMXBs also trace recent star formation activity (\mbox{$\tau < \textrm{a few} \times 10^7$ yr};~\citealt{Grimm03,Shtykovskiy05}).

As mentioned before, during the last few years models including the interaction of binary stars have been developed. Evolutionary population synthesis models by \cite{Cervinyo02} (hereafter, CMHK02) can reproduce satisfactorily the observed values of the soft X-ray luminosity in starbursts~\citep{Mas-Hesse08}. Also, CMHK02 models predict the order of magnitude of young binary systems which are active, i.e. actually emitting X-rays by accretion processes and therefore considered HMXBs, at a given evolutionary state of the burst and once an initial binary fraction, $f_{\mathrm{bin}}$, is assumed. Their predictions are thus sensitive to the existent binary fraction in a burst. We used the CMHK02 models to confirm the presence of HMXBs in those HII regions with detectable WR features and a counterpart in the Chandra image (see Fig.~\ref{fig:x_ray_im}, left). Thus, we estimated the order of magnitude of the number of active binaries in NGC 3310, as well as their integrated hard X-ray emission $L_{2-10~\mathrm{keV}
}$ assuming a typical luminosity \mbox{$L_\mathrm{X} \sim 10^{38}$ erg s$^{-1}$} for every single HMXB.

Excluding the nucleus, the age of the ionizing population of \hii regions hosting WRs ranges from 3 to about 5 Myr, and their integrated mass is of the order of $4\times10^7$ \msun\onespace. According to the CMHK02 models (for $Z = 0.4 Z_{\odot}$) and assuming a binary fraction of $f_{\mathrm{bin}} = 0.5$, this is consistent with about 100 active HXMB and an integrated hard X-ray luminosity of \mbox{$L_{2-10~\mathrm{keV}} \sim 10^{40}$ erg s$^{-1}$}. This prediction is similar within a factor of 2 to the observed $L_{2-10~\mathrm{keV}}$ value yielded by the fitting models of the XMM-Newton spectra already described above. Given the uncertainties inherent to this sort of estimations, and bearing in mind that CMHK02 models only predict the order of magnitude of the number of active binaries, we argue that there exists an excellent agreement between the predicted and the observed hard X-ray luminosity values for NGC 3310. If, on the other hand, no presence of binaries is considered in the CMHK02 models (i.
e., $f_{\mathrm{bin}} = 0$), the only contributions to the hard X-ray emission are those from the SNRs and the hard-energy tail of the  thermal contribution from the gas heated by the starburst activity. Under the same initial parameters of mass, age and metallicity, the integrated predicted hard X-ray emission is of the order of \mbox{$L_{2-10~\mathrm{keV}} \sim 5\times 10^{38}$ erg s$^{-1}$}, i.e., a factor of 20 lower than that predicted with $f_{\mathrm{bin}} = 0.5$, and even lower compared to the observed value using the \textit{XMM-Newton} data. In conclusion, the predicted X-ray luminosity using the CMHK02 models including binaries is close to the measured X-ray luminosity in circumnuclear \hii regions with strong presence of WRs. This result strengthens the notion discussed in the previous section that additional processes, such as the binary population, should be included in stellar population modelling if a correct characterization of the WR population is to be made.

\begin{figure*}
 \centering
\includegraphics[trim = 3.5cm 0cm 4cm -1cm,clip=true,angle=90,width=0.90\textwidth]{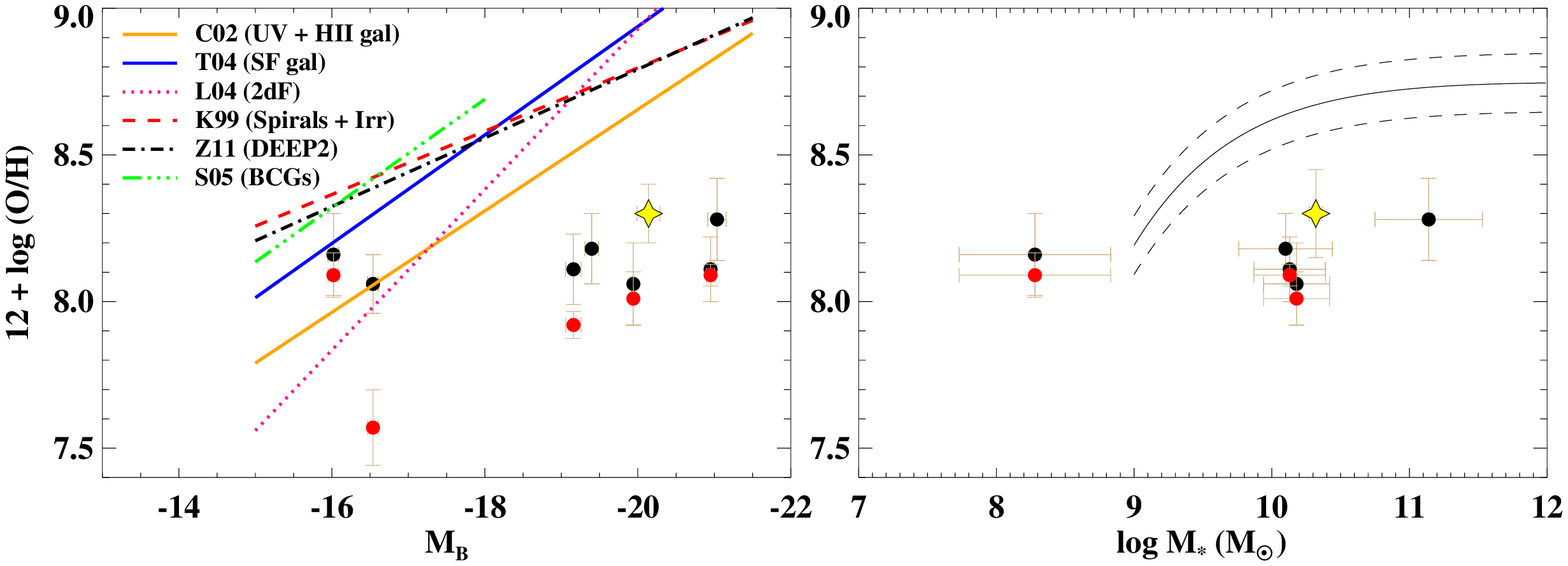}
  \caption{\textbf{Left:} The $L - Z$ relation of NGC 3310 (yellow star) and long-duration GRB hosts in H10, being in black the original values and in red our recalculated metallicities using the C-method (see text). Luminosity-metallicity relation for various galaxy samples are drawn from the literature: UV-selected galaxies from Contini et al.~(\citeyear{Contini02}) (solid orange line), SDSS star-forming galaxies from Tremonti et al.~(\citeyear{Tremonti04}) (solid blue line), a large magnitude-limited sample from Lamareille et al.~(\citeyear{Lamareille04}) (dotted pink line), irregular and spiral galaxies from Kobulnicky et al.~(\citeyear{Kobulnicky99a}) (dashed red line), emission-line galaxies at \mbox{z $\sim$ 8} from the Deep Extragalactic Evolutionary Probe 2 survey~(Zahid et al.~\citeyear{Zahid11}; dotted-dashed black line) and star-forming BCGs from Shi et al.~(\citeyear{Shi05}) (three dotted-dashed green line).~\textbf{Right:} The $M - Z$ relation of NGC 3310 (yellow star) and long-duration GRB 
hosts in 
H10 (in black and red, as explained before). The relation obtained for the mass-metallicity relation of a sample of 150 blue/star-forming galaxies observed with CALIFA (S\'anchez et al.~\citeyear{Sanchez13}) is overplotted as a solid line. Dashed lines indicate the 0.1 dex dispersion around Sanchez's $M - Z$ relation.}
  \label{fig:wr_metal_MB_mass}
\end{figure*}

\subsection{WR-GRB connection. The importance of the environment}

As we have just seen, the WR population can be well related to high energy processes. In this section, we investigate if this population is also behind the most energetic events in the Universe, the Gamma-ray bursts (GRBs). The mechanisms that can produce such amounts of photons with energies as high as several MeV and GeV and the identification of their progenitors remain disputed.

Short GRBs (with a duration of \mbox{$\leq$ 2 s} and hard spectrum) are believed to originate from the merger of compact binaries. On the contrary, long GRBs (with a duration \mbox{$\geq$ 2 s} and soft spectrum) are thought to result from the extreme gravitational collapse of a rapidly rotating, massive star (core-collapsar model;~\citealt{Hartmann05}, and references therein). The most favoured parent supernova (SN) population of GRBs is formed by peculiar type Ibc SNe. WRs are naturally considered to be the most favoured candidates of long duration GRB progenitors. According to models, a lower limit to the metallicity of sub-solar value (i.e., $Z \leq 0.2 - 0.4 Z_{\odot}$) is needed~\citep{Hirschi05}, though magnetic--field breaking present some difficulties in producing GRBs~\citep{Petrovic05}. WRs have indeed been observed in several GRB hosts (\citealt{Han10}; hereafter H10). Here we compare the properties of these hosts and the integrated properties of NGC 3310. A detailed study on the 
environment of a local galaxy with similar properties than GRB hosts is essential to better understand the physical properties of GRBs observed at moderate and high redshifts and the nature of their progenitors.

H10 performed a spectral analysis of 8 long-duration GRBs hosts in order to study the environment in which such energetic events can take place. Since, according to the core-collapsar model, WRs are considered as the most favored candidates of the progenitor of long-duration GRBs, the presence of WRs in their hosts provided evidence in favour. They identified other characteristics that support the core-collapsar model, such as the observed high WR/O star ratio (i.e., 0.01--0.20) and the low metallicity of their sample of GRB hosts. We do indeed find for the regions in NGC 3310 similar ratios. H10 studied the luminosity and stellar mass-metallicity relations to relate GRB hosts with the WR population. They 
represent fundamental relations which indicate the evolutionary status and star-formation histories of the galaxies. We have thus derived the integrated luminosity, mass and metallicity of NGC 3310 and have compared these relations with those obtained in H10.

First of all, the adopted oxygen abundance for NGC 3310 corresponds to \mbox{12 + log(O/H) = 8.25 $\pm$ 0.10} since, according to \mbox{Paper~I}, this is the typical abundance measured at one effective radius and in any case the metallicity gradient in NGC 3310 is rather flat. Note that for a galaxy with a steep metallicity gradient the characteristic oxygen abundance, representative of the average value across the galaxy, corresponds basically to that at one effective radius in the Local Universe~\citep{Zaritsky94,Sanchez13}. Then, we took the photometric $B$-band magnitude of this galaxy from the literature (in particular, from NED\footnote{http://ned.ipac.caltech.edu/}). 

In Fig.~\ref{fig:wr_metal_MB_mass} (left) we present the $L - Z$ diagram for GRB hosts showing detectable WR features in H10 and for NGC 3310. H10 made use of a calibration based on the $R_{23}$ parameter (\citealt{Pagel86}) to obtain abundances. Although calibrations based on this parameter are widely used, they are known to present important issues at moderately-low metallicities, just the metallicities derived in H10. Nevertheless, H10 provide the line fluxes for the different element species (i.e., \nii, \oiii, \sii), which has allowed us to recalculate such estimations. To that end and with the knowledge that the metallicity could in principle be lower than \mbox{12 + log(O/H) = 8.0} we have used the Counterpart method (C-method), described in~\cite{Pilyugin12b}. This method is based on the standard assumption that \hii regions with similar intensities of strong emission lines have similar physical properties and abundances. It basically selects a number of reference (well-measured abundances) \hii 
regions and then the abundances in the target \mbox{\hii} region are estimated through extra-/interpolation. Although all the figures with oxgyen abundance and N/O ratios (in next section) shown refer to the values obtained using the C-method, we have also derived independent values using the recently published HII-CHI-mistry~\citep{Perez-Montero14}, based on updated grids of photoionization models. Although differeces arise within 0.1--0.2 dex between both methods, the tendences described in this work are the same. 

As can be seen in the figure, the new abundance values are generally similar to those derived in H10, though for two cases the offset is \mbox{$\geq 0.2$ dex}, larger than the typical uncertainties of the calibrations used. The characteristic abundance of NGC 3310 is similar or somewhat higher than the characteristic abundance of H10's GRB hosts. As for the most luminous ones (including NGC 3310), the discrepancy with other samples from the literature (star-forming, spiral, irregular and compact dwarf galaxies) is evident. GRB hosts and NGC 3310 have lower metallicity values than other galaxies at similar luminosities. This trend is also discussed in~\cite{Levesque10} and~\cite{Graham13}. 

We have derived the stellar mass of NGC 3310 following the recipe given in~\cite{Bell03}, which relates the stellar mass-to-light ratio and the colour of a galaxy to its mass. In particular, we have used the $(B - V)$ colour and $K$-band absolute magnitude to estimate the stellar mass of NGC 3310, using the following expression:

\begin{equation}
 \mathrm{log}\left(\frac{M_{\star}}{M_{\odot}}\right) = -0.4 (M_K - 3.28) + [a_K + b_K(B - V) + 0.15],
\end{equation}
where the coeffitiens $a_K$ and $b_K$ are taken from Table 7 in~\cite{Bell03}. With this we plot the $M - Z$ relation for NGC 3310 and compare it with the one obtained for GRB hosts in H10 and the $M - Z$ relation explored in the CALIFA sample~\citep{Sanchez13} for 150 blue/star-forming galaxies (Fig.~\ref{fig:wr_metal_MB_mass}, right). Again, the discrepancy between NGC 3310 and GRB hosts with local star-forming galaxies (this case with comparable stellar masses) is obvious. 

\begin{figure*}
\centering
\includegraphics[trim = 3.0cm 0cm 3.5cm -1cm,clip=true,angle=90,width=0.85\textwidth]{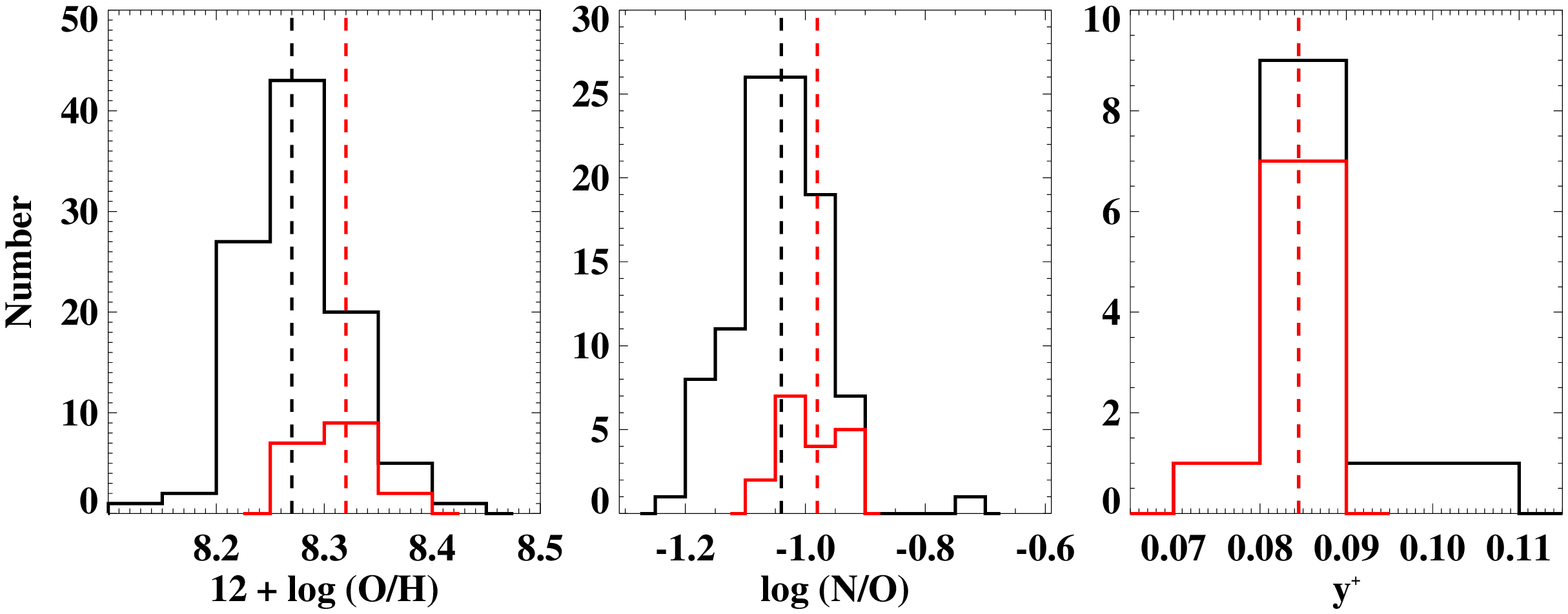}
  \caption{Histograms with the oxygen abundance (left), oxygen to nitrogen abundance ratio (centre) and $y^{+}$ (right) for the sample of \hii regions identified in \mbox{Paper~I} (black) and the sample of \hii regions with detected WR in this study (red). Vertical dashed lines mark the median value of each distribution.}
  \label{fig:chemical_enrichment1}
\end{figure*}

All in all, NGC 3310 share properties with GRB hosts located at higher redshifts and where the presence of WRs has been verified. These properties differ from those of normal local star-forming, compact and irregular galaxies. There is clear evidence that NGC 3310 collided with a poor-metal dwarf galaxy~\citep{Balick81,Schweizer88,Kregel01}, producing a minor merger event. Models of galaxy interactions predict that in a major merger event, radial mixing processes take place (e.g.,~\citealt{Barnes96,Rupke10a,Perez11}). As a consequence a pre-existing metallicity gradient flattens, as observed in several samples of interacting galaxies (e.g.,~\citealt{Chien07,Kewley10,Rupke10b,Rich12}), and more recently in a statistically significant sample of interacting and non-interacting galaxies in~\cite{Sanchez14}. The same scenario could happen in a minor merger event, as a few observations suggest~(\paperone;~\citealt{Werk11};~\citealt{Alonso-Herrero12}). It is not surprising that the typical abundance of the 
parent (prior to merger) galaxy has been lowered  due to metal mixing processes induced by the merger event. Therefore, a minor event with a poor-metal object 
seems to be the logical process to understand the discrepancies of some properties of NGC 3310 discussed here with those of other type of galaxies and starbursts. This induces us to speculate if GRB hosts are preferentially 
products of merger events, which can trigger star-formation and flatten previous abundance gradients of the gas. In fact, several studies support that nuclear starbursts in interacting galaxies (even for minor mergers) require a top-heavy initial mass function (IMF), preferentially producing high-mass stars~(e.g.~\citealt{Rieke80,Doyon92,Gibson97,Baugh05,Brassington07,Espinoza09,Bartko10,Habergham10,Habergham12}), although this is still a controversial issue (see~\citealt{Bastian10} for a recent review). If high-mass stars are preferentially produced, then more WRs should be expected in the central regions of galaxy interactions and past mergers. Thus, if WRs are progenitors of GRBs, they would be more likely to happen in these environments.

\subsection{Chemical enrichment by WR winds}

As it is clearly reflected in the previous sections, WRs are involved in energetic processes. These processes eject large amounts of metals to the interstellar medium (ISM). Not only the SN explosions but also the powerful stellar winds that the most massive O stars undergo before and during the WR phase can pollute the ISM with metals in relatively short time-scales. In fact, since early in the 90's it is well known that WRs are significant contributors of He, N and C to the ISM~\citep{Maeder92}. In this section we compare the abundance of different species so as to know the extent of the rapid chemical pollution caused primarily by the WRs.

In \paperone~we derived the oxygen abundance of the gas for the complete sample of \hii regions in NGC 3310. We compare in Fig.~\ref{fig:chemical_enrichment1} (left) the abundances for the whole sample with those for \hii regions hosting WRs. Although the distributions seem different, the difference in median values is less than 0.1 dex, larger than the typical systematics (i.e, $\sim$0.2 dex), and according to the Kolmogorov-Smirnov test (KS) the significance level at which both distributions can be considered different does not reach 99\%. Therefore, the oxygen abundance is roughly homogeneous in NGC 3310 over spatial scales of few hundreds of pc. This is in agreement with the flat abundance gradient reported in \mbox{Paper~I}.

\begin{figure*}
 \centering
\includegraphics[trim = -1cm 3cm -1cm 4cm,clip=true,width=0.85\textwidth]{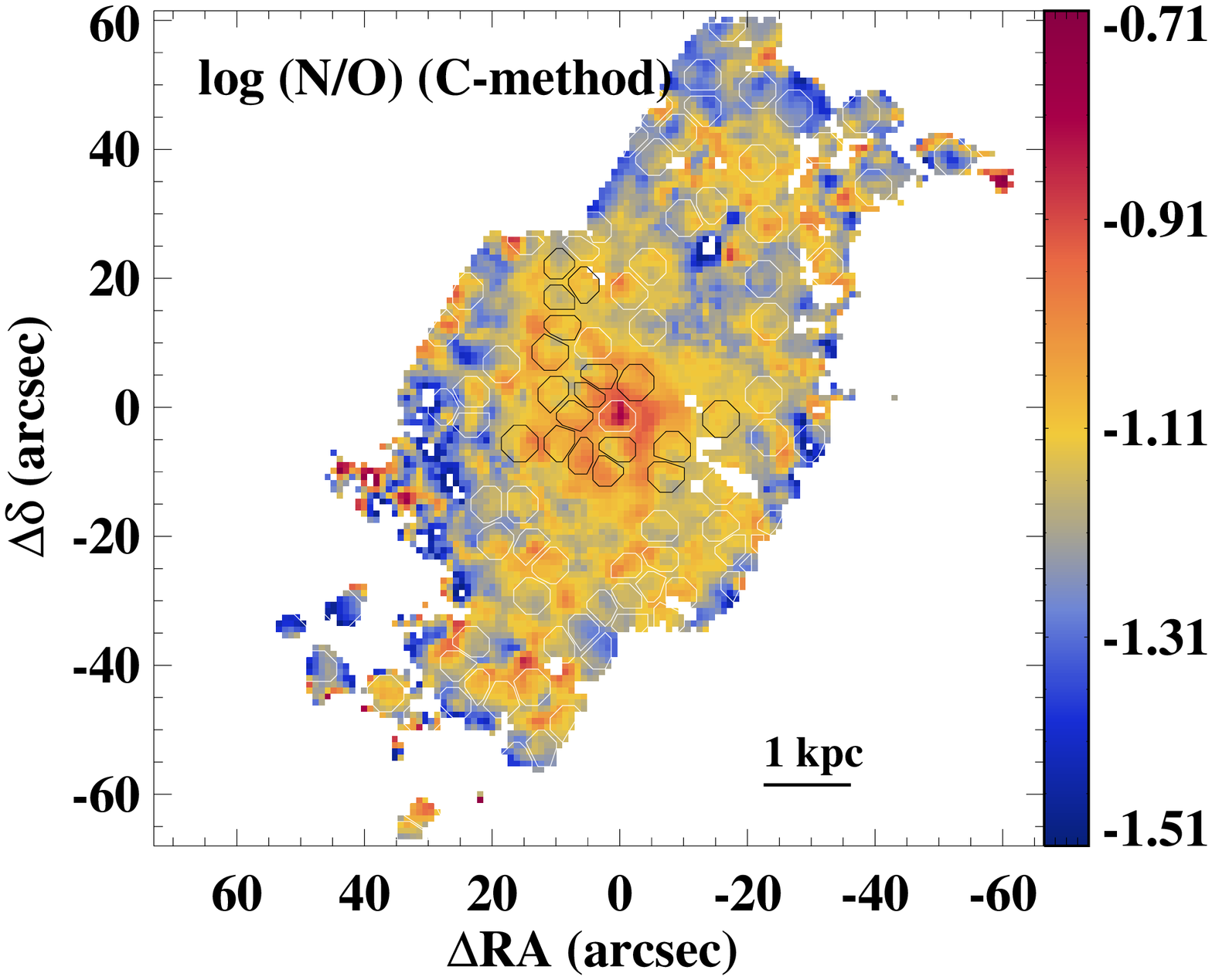}
  \caption{Map of the N/O ratio (obtained with the $O2N2$ parameter) with the apertures of the \hii regions overplotted. For those cases where detection of WR has been found, the apertures are drawn in black.}
  \label{fig:chemical_enrichment2}
\end{figure*}

A similar comparison has been done on the nitrogen to oxygen abundance ratio. With the methods described in the previous section we could also obtained log (N/O). As Fig.~\ref{fig:chemical_enrichment1} (centre) shows, the median value of this distribution is about -1.05 if we consider all identified \hii regions in NGC 3310. When we make the histogram only for \hii regions with measurable WR features, the histogram seems to be shifted 0.1 dex towards higher N/O ratios (i.e., higher N abundances). Although the difference in the median value of both distributions is just 0.06 dex, According to the Kolmogorov-Smirnov test (KS) the significance level at which both distributions can be considered different is higher than 99\%. Thus, our measurements suggest that pollution by WRs cannot be ruled out in some cases at scales of the order of 200--300 pc. We have explored in which cases this enrichment is more evident by inspecting the ratio in a spaxel--by--spaxel basis (see Fig.~\ref{fig:chemical_enrichment2}). 
The enhancement of the N/O ratio is more evident in the central regions, with orange colours in the maps indicative of ratios of the order of $\sim -0.9$. For the rest of \hii regions the yellow-greyish colours indicate ratios of the order of \mbox{[-1.1,-1.2]}. Here, each spaxel spans a linear distance of about 78 pc. Although a dithered pattern was followed when observations were taken in order to sample the space between fibres (with a diameter of 2.6\arcsec) and an interpolation and reconstruction was made afterwards so as to have a spatial binning of 1\arcsec per spaxel, with our data we cannot really investigate spatial variations below the spatial resolution of the data, about 200 pc.   

We can further assess the hypothesis of N enrichment taking into account that, if it comes from WR stellar winds, He enhancement is also expected (e.g.,~\citealt{Pagel86,Esteban92,Kobulnicky96}). The well detected and measured HeI $\lambda\lambda\lambda$ 4471,5876 and 6678 \AA{} lines were considered to calculate the singly ionized  helium abundances (\mbox{$y^+ \equiv$ He$^+$/H$^+$}). He line measurements at wavelengths 3889 and 4026 \AA{} were not considered since, given the spectral resolution of the data (\mbox{$\sim 10$ \AA{}/pixel}), the former is blended with the H8 Balmer line (at about 3889 \AA{}) and the latter is observed (\mbox{$\lambda_{\mathrm{obs}} \sim 4040$ \AA{}}) at a very close wavelength to the \mbox{HgI $\lambda$4046 \AA{}} sky line. The singly ionized helium abundance was calculated using the prescriptions provided by~\cite{Monreal13}. We did not apply any correction for fluorescence (two of them have a small dependence with optical depth effects but the \hii regions have low densities)
. We derived three singly ionized abundances. However, we only took the weighted average of two of them (those obtained via the $\lambda\lambda$5876,6678 lines) as the adopted value, since important discrepancies were sometimes observed with the value obtained via the $\lambda$4471 line, which was the weakest of the three and extremely dependedent on the applied absorption correction derived with \starlight. 

The histogram of $y^{+}$ for the whole sample of \hii regions and for the sub-sample with WR features is essentially the same (the KS probability that the distributions are the same is higher than 90\%). Under the assumption that the element abundances found in ring nebulae around Galactic WRs measured by~\cite{Esteban92} are reasonably representative of the heavy element yields of WR winds,~\cite{Brinchmann08a} estimated the expected average increase in log He/H and in log N/O in WR ejecta. The former estimation, of the order of 0.18 dex, is considerably lower than the latter, being \mbox{$\sim 0.85$} dex. Therefore, a higher sensitivity to changes in N/O are expected compared to He/H. Given the weak pollution of N that we have observed at scales of about 200 pc, the absence of pollution of He is consistent with the expectations.

Based on these results, although we did not robustly prove that WR stellar winds contribute significantly to the N enrichment at scales \mbox{$\gtrsim$ 200 pc}, in some cases this hypothesis cannot be simply ruled out. In fact, in this study we are close to the spatial limit at which the hot gas is chemically homogeneous. According to several studies, on spatial scales of about a few hundreds of pc, the ionized gas observed in \hii regions is chemically homogeneous (e.g.,~\citealt{Kehrig08,Perez-Montero09,Cairos09,Perez-Montero11}). On smaller scales and down to about 10 pc, chemical pollution has been previously reported for Galactic WR nebulae~\citep{Esteban92,Fernandez-Martin12} and for irregular and WR galaxies (e.g.,~\citealt{Lopez-Sanchez10b,Lopez-Sanchez11,Monreal12,Kehrig13,Perez-Montero13}). 

The processes that rule metal dispersal and mixing on the ISM are not well understood yet within time-scales of \mbox{$<$ 10 Myr} or \mbox{$\sim$ 100 Myr}. The fate of the metals released by massive stars in \hii regions is still an open question, and the processes beneath at work are hard to model~\citep{Recchi13a,Recchi13b}. To better understand how the ejected metals cool and mix with the ISM, further investigation on the metal content of the different ISM phases and at lower spatial scales is needed. Detection of any chemical pollution on scales of less than a hundred parsecs escapes this work because of our resolution element size.

\section[]{Conclusions}

We have analysed in detail the broad stellar features originated by winds of Wolf-Rayet stars detected in a sample of \hii regions identified in the distorted spiral galaxy NGC 3310. The use of the IFS technique has allowed us to study the spatial distribution of star-forming regions hosting WRs, characterize their WR content and set constraints on evolutionary synthesis models. We have also investigated the influence of these stars on the environment. The most important results of this study can be summarized as follows:

\begin{enumerate}[-]
 \item We have identified a total of 18 \hii regions distributed in the central regions of NGC 3310 and along the northern spiral arm. Interestingly, a few regions with similar derived ionized mass and lower continuum on the disc at larger galactocentric distances do not show clear evidence of stellar emission from WRs. We have performed a detailed fitting of the spectra of these regions considering the broad stellar and narrow nebular emission lines in the WR blue bump. Most of the broad band emission is likely to be originated from WNL stars, though a contribution of up to 10--20 \% from Carbon-type WRs cannot be discarded. Under these considerations and assuming metallicity-dependent luminosities for the WR features, we conclude that the regions host from dozens to a few hundreds of WRs, being their integrated number in \mbox{NGC 3310} somewhat larger than 4000.

 \item We have compared the number of WR to O stars ratio in our sample with empirical calibrations of this ratio using the intensity of the \heii $\lambda$4686 broad line or the $\lambda$4650 blend to derive the number of WRs. There is agreement within 0.1-0.2 dex only when correction of these ratios by dust absorption of UV photons by dust grains in the nebula is taken into account.
 
 \item Stellar synthesis models that do not include binaries or rotation in their prescriptions generally underpredict the observational measurements, i.e., EW (broad  $\lambda$4686), I(broad $\lambda$4686)/I(\hb\onespace). In most cases disagreement larger than factors of 2--3 occurs. Nevertheless, if a high fraction of binaries is included in models and photon leakage is taken into account, then observations and predictions come to a much better agreement. 

 \item The integrated hard X-ray luminosity (mainly originated in High Mass X-ray Binaries) of the central regions of NGC 3310 is of the order of \mbox{$L_{2-10~\mathrm{keV}} \sim 2.2 \times 10^{40}$ erg s$^{-1}$}. There is a spatial correlation between the X-ray sources and the \hii regions hosting WRs. With the knowledge of the age and mass of the ionizing population in these regions we have estimated that the binary fraction must be significant, which reinforces the necessity of including this scenario in the models. 

 \item We have discussed on the plausability that WRs can be progenitors to long-durantion Gamma-ray bursts (GRBs). We have found some similarities between integrated properties of NGC 3310 and a small sample of Gamma-Ray Burst (GRB) hosts for which the presence of WR emission with high N/O ratios has been observed. In particular, they all seem not to follow the well-known luminosity-metallicity and mass-metallicty correlations that have been observed in a variety of star-forming objects (e.g., UV-selected galaxies, SDSS star-forming galaxies, irregular and spiral galaxies), showing lower metallicity at a given luminosity. A past galaxy interaction can be behind the discrepancy with the mentioned correlations for NGC 3310. Therefore, the environment may play an important role in the mechanisms that induce GRBs.

 \item The nebular chemical abundance in these regions is in general homogeneous over spatial scales of 200--300 pc. However, in a few cases a weak evidence for N metal enrichment due to strong WR stellar winds has been found at scales close to 200 pc. Therefore, at lower scales WRs are likely to be able to affect the environment at relatively short time-scales.

\end{enumerate}

\label{sec:conclusions}

\section*{Acknowledgments}
We first express our gratitude to the anonymous referee for his careful revision, which forced us to check and revise our measurements.\\
This work has been partially supported by projects AYA2010-21887-C04-03 and AYA2010-21887-C04-01 of the Spanish National Plan for Astronomy and Astrophysics, and by the project AstroMadrid, funded by the Comunidad de Madrid government under grant CAM S2009/ESP-1496, partially using funds from the EU FEDER programme. It is also partially funded by the exchange programme `Study of Emission-Line Galaxies with Integral-Field Spectroscopy' (SELGIFS, FP7-PEOPLE-2013-IRSES-612701), funded by the EU through the IRSES scheme. HOF is funded by a postdoctoral UNAM grant. \\
We thank Miguel Cervi\~no for his thorough explanations on the description, use and interpretation of evolutionary population synthesis models.\\
DMC acknowledges John Eldridge for providing with updated modelled nebular emission line fluxes using the \bpass models.\\ 
FFRO acknowledges the Mexican National Council for Science and Technology (CONACYT) for financial support under the programme Estancias Posdoctorales y Sab{\'a}ticas al Extranjero para la Consolidaci{\'o}n de Grupos de Investigaci{\'o}n, 2010-2012. 
S.F.S acknowledges the {\it Plan Nacional de Investigaci\'on y Desarrollo} funding programmes, AYA2012-31935, of the Spanish {\it Ministerio de Econom\'\i a y Competitividad}, for the support given to this project.
This research has made use of observations obtained with \mbox{$XMM-Newton$}, an ESA science mission with instruments and contributions directly funded by ESA Member States and NASA; and of data obtained from the Chandra Data Archive.

\bibliographystyle{mn2e}

 \bibliography{my_bib.bib}{}

\clearpage

\appendix

\section{Appendix}

\noindent\begin{minipage}{\textwidth}
\hspace{1cm}
 \includegraphics[trim = -1cm -1cm 0cm -1cm,clip=true,width=0.85\textwidth]{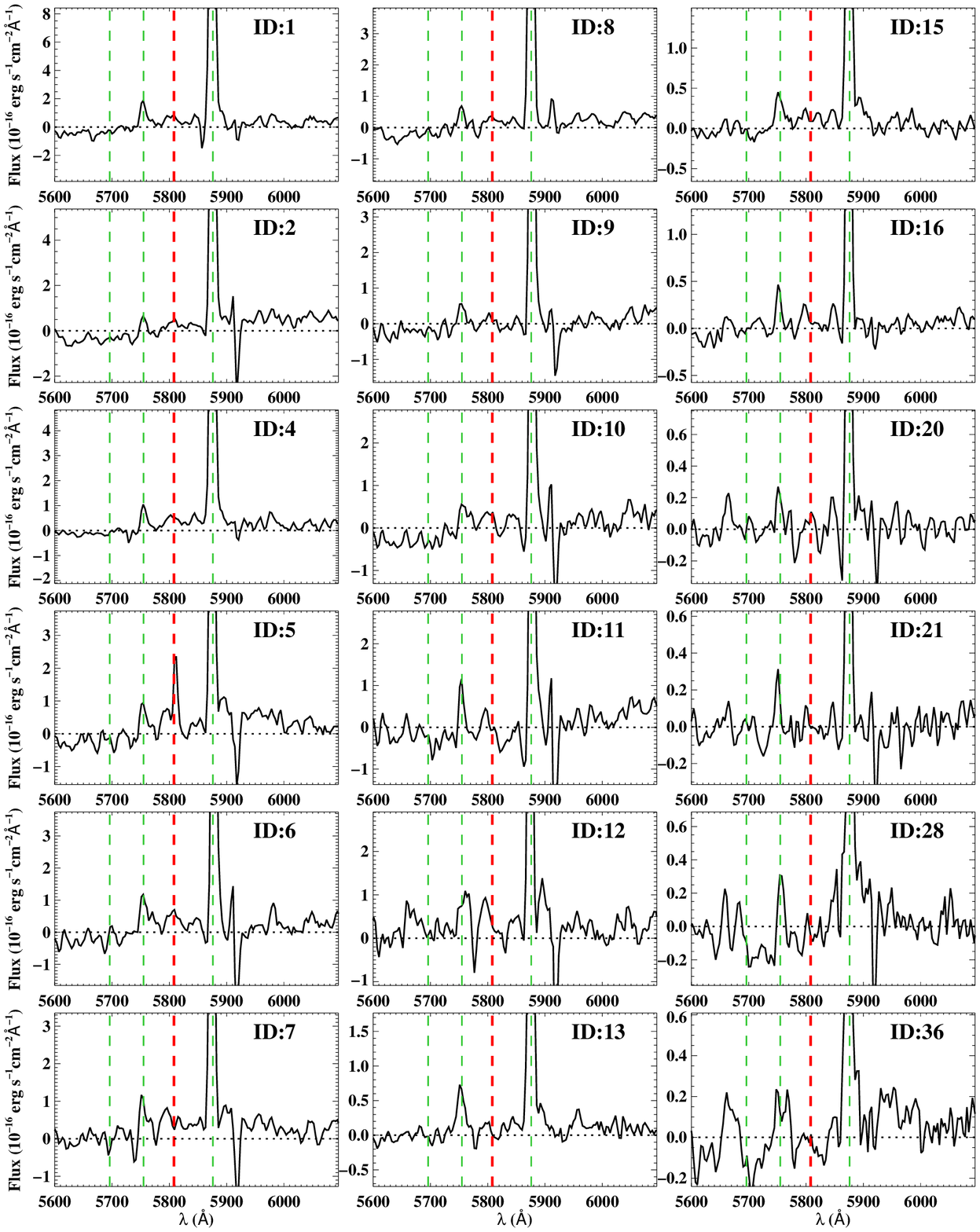}
  \captionof{figure}{Spectra of the 18 identified \hii regions with detected blue bump in a window close to where the red bump is expected (vertical red line). From left to right, vertical green lines, refer to the \ciii line at \mbox{5696 \AA{}} (WC or WO are the dominant contributors), the \nii~line at \mbox{5755 \AA{}} and the \hei~line at \mbox{5876 \AA{}}.}
 \label{fig:red_bump}
\end{minipage}


\label{lastpage}

\end{document}